\documentclass[12pt]{article}  
\usepackage[english]{babel}   %[english]

\usepackage[T1]{fontenc}
\usepackage{graphicx}     

\usepackage{times}

\usepackage{amsmath}
\usepackage{amsfonts}
\usepackage{amssymb}

\newcommand{\D}{{\rm d}}

\newcommand{\realp}{{\mathbb{R}}{\rm e}\,}
\newcommand{\imp}{{\mathbb{I}}{\rm m}\,}

\newtheorem{theorem}{Theorem}

\newcounter{llista}
\begin{document}
\title{Causality and dielectric functions \\for linear media with spatial dispersion}
%Kramers-Kronig relations \\ beyond the optical approximation}
\author{Josep Llosa\thanks{e-mail address: pitu.llosa@ub.edu} \;and Francesc Salvat\thanks{e-mail address:  cesc@fqa.ub.edu}\\[1ex]
%\affiliation{
Facultat de F\'{\i}sica (FQA and ICC), Universitat de Barcelona\\  Diagonal 645, 08028 Barcelona, Catalonia, Spain } 
\maketitle

\begin{abstract}
We extend Kramers-Kronig relations beyond the optical approximation to dielectric functions  
that depend not only on frequency but on the wave number as well. This implies extending the notion of causality commonly used in the theory of Kramers-Kronig relations to include the fact that signals cannot propagate faster than light in vacuo. 
The extension is applied to some microscopic models for the dielectric function and is compared with previous generalizations.\\
The results derived here also apply to general theories of isotropic linear response in which the response function depends on both wave number and frequency.
\end{abstract}

\section{Introduction  \label{S0}}
% Kramers-Kronig PRB v09  %%% Hi ha la teva disertació que, a la millor, ha de canviar de to.

%%%%%%%%%%%%%%%%%%%%%%%%%%%%%%%%%%%%%%%%%%%%%%%%%%%%%%%%%%%
Kramers-Kronig relations are restrictions \cite{Jackson}, \cite{Zangwill} that apply to the complex, frequency dependent optical dielectric function  $\varepsilon(\omega)$ ---or to the electric susceptibility $\chi(\omega)\,$--- of a linear material medium in the optical approximation, when the dependence on the wave vector $\mathbf{q}$ is neglected, i. e.  the applied fields do not vary appreciably in space. Such relations follow from two rather general assumptions, namely (1) causality in time and (2) a certain high-$\omega$ behaviour of the optical dielectric function (ODF).

%%%%%%%%%%%%%%%%%%%%%%   Resposta lineal   %%%%%%%%%%%%%%%%%%%%%%%%%%%%%%%%%%%%%%%%%%%%%%%%%%%%%%%%%%%%%%%%%%%%%%%%%%%%
They generally apply in the context of any theory of linear response \cite{Nussenzveig1972} based on a triple, Input-Output-Response function, that are connected by a relation of the sort 
\begin{equation}  \label{0a}
 \mathbf{P}(\mathbf{x},\omega) = \chi(\omega) \, \mathbf{E}(\mathbf{x},\omega)    \,, 
\end{equation}
In this particular instance the triple is Electric field-Polarization-Electric Susceptibility. A similar connection is found in scattering processes, e. g. \cite{Schutzer1951} and \cite{VanKampen1953}, where the triple is made of Incoming wave, Scattered wave and $S$-matrix.

%%%%%%%%%%%%%%%%%%%%%%%%%%%%%%%%%%%%%%%%%%%%%%%%%%%
The interest of Kramers-Kronig relations is twofold: first, they provide a useful tool to analyze the consistency of empirical ODFs and, second, they are also a quality test that has to be passed by any acceptable expression of $\varepsilon(\omega)$ derived from whichever microscopic model, e. g. \cite{Lindhard1954}.

To study the implications of causality for fields varying in both time and space it is necessary to account for the effect of retardation,  i. e. for the finite velocity of propagation of the fields. The goal of the present work is to derive a generalization of the Kramers-Kronig  relations for the entire dielectric function (DF), that is  for $q\ne 0$. Aside from its fundamental interest, this generalization can be employed to exhibit the consistency, or the lack of it, of extension algorithms adopted in optical-data models. 
%%%%%%%%%%%%%%%%%%%%%%%%%%%%%%%%%%%%%%%%%%%%%%%%%%%%%%%%%%%%%%%%%%%

In Kramers-Kr\"onig relations causality is understood as
\begin{quote}
The polarization vector $\mathbf{P}(\mathbf{x},t)$ only depends on the values of the electric field $\mathbf{E}(\mathbf{x},t^\prime)$ at the same place and at previous instants of time, $t^\prime \leq t$.
\end{quote}

If a linear isotropic medium responds to an applied harmonic, or monochromatic, electric field $\mathbf{E}(\mathbf{x},\omega)\,e^{-i\omega t}$ by acquiring a polarization $\mathbf{P}(\mathbf{x},\omega)\,e^{-i\omega t}$, both magnitudes are proportional, whence the relation (\ref{0a}) follows,
%\begin{equation}  \label{0a}
%  \mathbf{P}(\mathbf{x},\omega) = \chi(\omega) \, \mathbf{E}(\mathbf{x},\omega)   
%\end{equation}
with the factor  $\chi(\omega)$ depending on the frequency. 
Linearity also implies that the response $\mathbf{P}(\mathbf{x},t)$ to a superposition of harmonic fields 
$\quad\displaystyle{  \mathbf{E}(\mathbf{x},t) = \frac1{\sqrt{2 \pi}} \,\int_\mathbb{R} \D\omega\,e^{-i\omega t} \,\mathbf{E}(\mathbf{x},\omega) }\,$ 
is the superposition of responses
%$$\mathbf{P}(\mathbf{x},t) = \frac1{\sqrt{2 \pi}}\,\int_\mathbb{R} \D\omega\,e^{-i\omega t} \,\chi(\omega)\,\mathbf{E}(\mathbf{x},\omega) $$ 
and, by the convolution theorem \cite{Vladimirov}, 
\begin{equation}  \label{0b}
 \mathbf{P}(\mathbf{x},t) =  \frac1{\sqrt{2 \pi}}\,\int_{\mathbf{R}} \D t^\prime \, \tilde\chi(t-t^\prime) \,\mathbf{E}(\mathbf{x},t^\prime) \,, \qquad  {\rm with} \quad \tilde\chi(t) = \frac1{\sqrt{2 \pi}} \,\int_\mathbb{R} \D\omega\,e^{-i\omega t} \,\chi(\omega) 
\end{equation}

Causality means that the {\em effect} $\mathbf{P}(\mathbf{x},t)$ cannot depend on the {\em causes} $\mathbf{E}(\mathbf{x},t^\prime)$ at later times, which implies that
\begin{equation}  \label{0c}
 \tilde\chi( t - t^\prime) = 0 \qquad {\rm for} \qquad t^\prime > t \,.
\end{equation}
We shall refer to this as {\em local causality} (L-causality), to distinguish it from other causality conditions that will be introduced below.

Often the optical approximation, does not suffice to account for the experimental results. The present view of a material medium as linearly reacting to the excitation produced by the electromagnetic field is a macroscopic description which can be modeled as the linear approximation to the collective response of the elementary charges contained in the medium. The different microscopic models \cite{Lindhard1954} basically consist of analyzing the energy and momentum exchange among field and charges. These exchanges occur in multiples of $\hbar\omega$ and $\hbar\mathbf{q}$, respectively. In some instances ---for ``small'' $|\mathbf{q}|$--- the optical approximation $\varepsilon(\omega)$ yields a fairly good description of experimental results, but for higher values of  $|\mathbf{q}|$, the effect of momentum exchange becomes important and it is necessary to consider the dependence on the wave vector as well.

If we go beyond the optical approximation, the dielectric function $\varepsilon = 1 + 4 \pi\,\chi\,$ depends on the frequency and on the wave vector of the electric field and the same holds for electrical susceptibility. 
%$$\chi(\mathbf{q},\omega) = \frac{\varepsilon(\mathbf{q},\omega) - 1}{4 \pi}\,. $$ 
As a consequence the causality condition must be more involved. Indeed, an analogous expression to (\ref{0a}) for the polarization resulting from a linear isotropic response to a plane monochromatic wave is\footnote{In fact, the relation is a little less simple and susceptibility is a matrix rather than a scalar. The consequences of this are discussed in Appendix A.}
\begin{equation}   \label{e2}
 \mathbf{P}(\mathbf{q},\omega)\,e^{i(\mathbf{q}\cdot\mathbf{x} - \omega t)} = \chi(\mathbf{q},\omega) \,\mathbf{E}(\mathbf{q},\omega) \,e^{i(\mathbf{q}\cdot\mathbf{x} - \omega t)}  
\end{equation}
As before, linearity implies that the response $ \mathbf{P}(\mathbf{x},t)$ to a superposition of plane monochromatic electromagnetic waves, 
$$\, \mathbf{E}(\mathbf{x},t) = \frac1{(2\pi)^2}\,\int_{\mathbb{R}^4} \D \mathbf{q}\,\D\omega\, \mathbf{E}(\mathbf{q},\omega) \, e^{i(\mathbf{q}\cdot\mathbf{x} - \omega t)} \, $$  
is the superposition of responses 
%given by
%$$\, \mathbf{P}(\mathbf{x},t) = \frac1{(2\pi)^2}\,\int_{\mathbb{R}^4} \D \mathbf{q}\,\D\omega\,\chi(\mathbf{q},\omega) \mathbf{E}(\mathbf{q},\omega) \, e^{i(\mathbf{q}\cdot\mathbf{x} - \omega t)} $$
and, by the convolution theorem for Fourier transforms, we have
\begin{equation}   \label{e1}
\mathbf{P}(\mathbf{x},t) =  (2\pi)^{-2}\,\int_{\mathbf{R}^4}  \D \mathbf{y}\,\D t^\prime \, \tilde\chi(\mathbf{x} - \mathbf{y},t -t^\prime) \,\mathbf{E}(\mathbf{y},t^\prime)   \,, 
\end{equation}
where  
\begin{equation}   \label{e3}
\tilde\chi(\mathbf{x},t)= (2\pi)^{-2}\,\int_{\mathbb{R}^4} \D \mathbf{q}\,\D\omega\,\chi(\mathbf{q},\omega) \, 
e^{i(\mathbf{q}\cdot\mathbf{x} - \omega t)}
\end{equation}
is the susceptibility in spacetime variables. In what follows, we shall use the same letter to indicate a physical magnitude, either as a function in wave vector-frequency variables $(\mathbf{q},\omega) $ or in spacetime variables $(\mathbf{x},t)$. In this second case the symbol is indicated with the diacritic ``$\;\tilde{ }\;$''.

Notice that equation (\ref{e1}) is invariant by spacetime translations, i. e. replacing the electric field by $\mathbf{E}^\prime(\mathbf{y},t^\prime)= \mathbf{E}(\mathbf{y} - \mathbf{x}_0,t^\prime - t_0)\,$ results in a new polarization $\mathbf{P}^\prime(\mathbf{y},t^\prime)= \mathbf{P}(\mathbf{y} - \mathbf{x}_0,t^\prime - t_0)\,$. Thus the convolution relation (\ref{e1}) implies that the material medium is homogeneous, whence it follows that assuming the relation (\ref{e2}) we actually assume homogeneity. 

In turn the susceptibility in momentum space is the inverse Fourier transform
$$  \chi(\mathbf{q},\omega)= (2\pi)^{-2}\,\int_{\mathbb{R}^4} \D \mathbf{x}\,\D t\, \tilde\chi(\mathbf{x},t)\, 
e^{i(\omega t - \mathbf{q}\cdot\mathbf{x} )}  $$

The expression (\ref{e1}) can be understood as though the polarization at point $\mathbf{x}$ at instant $t$ is the ``effect'' of infinitely many ``causes'', namely the values of the electric field at every place $\mathbf{y}$ and every instant $t^\prime$. We then expect that the influence of $\mathbf{E}(\mathbf{y},t^\prime)$ on $\mathbf{P}(\mathbf{x},t)$ does not travel faster than light in vacuum. In terms of spacetime variables this  causality condition  reads
\begin{quote}
{\em The polarization $\mathbf{P}(\mathbf{x},t)$ only depends on the values of the electric field $\mathbf{E}(\mathbf{y},t^\prime)$ in the {\em absolute past}, i. e. such that the event $(\mathbf{y},t^\prime)$ is in the past light cone with vertex $(\mathbf{x},t)$, or 
$\; 0 \leq |\mathbf{x}-\mathbf{y}| \leq t - t^\prime  \,$  }
\end{quote}
\noindent
(we use natural units and take $c=1$\,).
We will thus refer to this condition as {\em finite speed causality} (FS-causality) to distinguish it from  {\em instantaneous causality} (I-causality), which might involve signals propagating as fast as necessary so as to permit an event at $t$ to be influenced by any event at $t^\prime \leq t$ no matter what the distance separating both. 

%\noindent
%\fbox{
%\parbox{17cm}{
Notice that FS-causality includes L-causality as a particular case. Under FS-causality the response function $\tilde\chi(\mathbf{x},t)$ is subject to the condition
\begin{equation}   \label{e4}
 t < |\mathbf{x}| \qquad \Rightarrow \qquad  \tilde\chi(\mathbf{x},t) = 0 
\end{equation}
and therefore $\quad   t < 0 \quad \Rightarrow \quad  t < |\mathbf{x}| \quad \Rightarrow \quad  \tilde\chi(t) = \tilde\chi(\mathbf{x},t) = 0 \,$.
This is due to the fact that in the optical approximation the convolution formula (\ref{0b}) involves only one space point, hence it does not imply the propagation at a distance of any signal.
%} }

%%%%%%%%%%%%%%%%%%%%%%   Resposta lineal   %%%%%%%%%%%%%%%%%%%%%%%%%%%%%%%%%%%%%%%%%%%%%%%%%%%%%%%%%%%%%%%%%%%%%%%%%%%%
In the same way as Kramers-Kr\"onig relations, our approach can be also applied in the context of any theory of linear response \cite{Nussenzveig1972} based on a triple, Input-Output-Response function, that are connected by a relation of the sort (\ref{e1}) or (\ref{e2}), where the triple is Electric field-Polarization-Susceptibility. A similar connection is found in scattering processes, e. g. \cite{Schutzer1951} and \cite{VanKampen1953}, where the triple is made of Incoming wave, Scattered wave and $S$-matrix.

The condition that we have called FS-causality has been explicitly invoked elsewhere, see refs. \cite{Toll1956} to \cite{Kirzhnits1987} to quote a few but, curiously, Toll's paper \cite{Toll1956} obviates the spatially dispersive case, in which the response function should depend also on the wave vector $\mathbf{q}\,$. 

Leontovich's work \cite{Leontovich1961} is the most accomplished previous attempt to extend the Kramers-Kr\"onig relations to media with spatial dispersion. He considers the connection of electric field (input) and current density (output) through conductivity $\sigma(q,\omega)$ (response function) and assumes that the latter is the same in all inertial reference frames. His study is restricted to one space coordinate and it is not clear whether his reasoning applies to higher dimensions. Moreover, the assumption that $\sigma$ is a Lorentz scalar is criticized by some authors ---see  \cite{Melrose1977} and \cite{SunPuri1989}--- who, on relaxing this requirement, arrive at the same relation as Leontovich. This suggests that Lorentz invariance or covariance of $\sigma$ is not essential for Leontovich's generalization and  we shall here derive it on the basis of more general assumptions. 

In what follows we will work out the implications of finite speed causality on the electric susceptibility $\chi(q,\omega)\,$ (we will be restricted to isotropic media). We shall proceed in much the same way as standard textbooks \cite{Jackson} derive Kramers-Kronig relations. In section \ref{S1} we prove that FS-causality implies that susceptibility is a doubly analytic function in some region in $\mathbb{C}^2$ and it has consequences on its asymptotic behavior as well. In Section \ref{S4} we derive a generalization of Kramers-Kronig relations suitable for FS-causality and compare it with previous results \cite{Leontovich1961}. In Section \ref{S5} several dielectric functions commonly used in the literature are examined to check whether the FS-causality condition is satisfied or not. 
Finally Section \ref{S6} is devoted to discuss and compare our generalization of Kramers-Kr\"onig relations with previous proposals, \cite{Leontovich1961}, \cite{Melrose1977}, \cite{Kirzhnits1987}, \cite{SunPuri1989} and \cite{Dolgov1982}.

\section{Consequences of finite speed causality on the susceptibility function \label{S1}}
In the present work we shall focus on isotropic media, i. e. $\,\tilde\chi(\mathbf{x},t)\ = \tilde\chi(r,t)\,$  only depends on the distance $\,r=|\mathbf{x}|\,$ and not on the direction of $\,\mathbf{x}\,$, and therefore  $\,\chi(\mathbf{q},\omega)\ = \chi(q,\omega)\,$\footnote{Isotropy is actually a more elaborated condition if the relation connecting polarization and electric field is of tensor type instead of merely the scaling  (\ref{0a}). We will comment it further in Appendix A. }.

The above mentioned FS-causality condition specifically requires that $ \,\tilde\chi (\mathbf{x} - \mathbf{y},t - t^\prime) \,$  be different from zero only if $\,0 \leq |\mathbf{x}-\mathbf{y}| \leq  t - t^\prime  \,$, which implies that
%\begin{equation}   \label{e4}
%\tilde\chi (\mathbf{r},t) \propto \Theta(c t - r) \,, \qquad {\rm for} \qquad  r \geq 0
%\end{equation}
%where $r = |\mathbf{r}|$ and $\Theta\,$ is the Heaviside unit step function.
the susceptibility function in wave vector-frequency variables is
$$   \chi(\mathbf{q}, \omega) = \frac{i}{2\pi \,q}\, \int_0^\infty \D t\,\int_{-t}^{t}\D r\, r\,\tilde\chi(r,t) \, e^{i(\omega t - q r)}   \,, $$
where $r = |\mathbf{r}|$ and integration over the spherical coordinates $\theta$ and $\varphi\,$ has been performed. We have also extended $\,\tilde\chi\,$ to negative values of $\,r$ as 
\begin{equation}   \label{e4a}
\tilde\chi(-r,t) = \tilde\chi(r,t)\,,\qquad \qquad r > 0\,. 
\end{equation} 
Thus $\,\chi(\mathbf{q}, \omega)\,$ depends only on $\,\omega\,$ and  $\, q = |\mathbf{q}|\,$ as expected and, writing $\, \chi(q,\omega)\, $ to stress this fact, we have that
\begin{equation}   \label{e5}
q\,  \chi(q,\omega) =  \frac{i}{2\pi }\, \int_0^\infty \D t\,\int_{-t}^{t} \D r\, e^{i(\omega t - q r)}\, r \,\tilde\chi(r,t) \,,
\end{equation}
whence it  obviously follows that $\,\chi(q,\omega) \,$ is an even function of $\,q\,$.

Introducing now the light-cone variables 
\begin{equation}  \label{e5a}
  u^{\pm} = \frac12\,(t \mp r)\,, \qquad \qquad  k_{\pm} = \omega \pm q 
\end{equation}
and the  inverse relations
\begin{equation}  \label{e5b}
 t = u^+ + u^- \,,\qquad r = u^- - u^+ \,, \qquad \omega = \frac{k_+ + k_-}2 \,,\qquad q = \frac{k_+ - k_-}2 \,,
\end{equation}
we have that the ``volume elements'' transform as	
$\; \D t\,\D r = 2 \, \D u^+\,\D u^- \;$ and $\; 2\,\D q\,\D \omega = \D k_+\,\D k_- 	\;$, and the integration domain $\,\mathcal{D} = \left\{ (r,t)\,, \;| r |\leq t \right\} \, $ becomes the first quadrant $\,\mathcal{D}^\prime=[0, \infty)\times[0,\infty)\, $. 
We shall refer to the  variables $k_\pm$ as {\em light-cone frequencies} 
because, similarly as $\mathbf{q}$ and $\omega$ are the Fourier conjugates of $\mathbf{r}$ and $t$, $\,k_\pm\,$ are the Fourier conjugates of the light-cone variables $\,u^\pm\,$.

Using these variables and including the fact that $\,  \omega t - q r = k_+ u^+ + k_- u^- \,$, the relation (\ref{e5}) can be written as
\begin{equation}   \label{e6}
(k_+ - k_-)\, \xi(k_+,k_-) =  - \frac{2\,i}{\pi}\, \int_0^\infty \D u^+ \, e^{-(0-i k_+) u^+} \,\int_0^\infty \D u^- \, e^{-(0-i k_-) u^-}
\,(u^+ - u^-)\,\tilde\xi(u^+,u^-)\,,
\end{equation}
For the sake of future clarity we have written 
\begin{equation}   \label{e6z}
\xi(k_+,k_-) := \chi(q,\omega)\qquad {\rm   and} \qquad \tilde\xi(u^+,u^-) := \tilde\chi(r,t)\,, 
\end{equation}
that is we use a different symbol for the susceptibility depending on whether the independent variables are $(q,\omega)$ or  $(k_+,k_-) \,$. 
The expressions $\,0-i\,k_\pm\,$ in the exponents mean that the integrals are to be calculated for $\,\delta -i\,k_\pm\,$ and then take the limit for $\delta\rightarrow 0^+\,$ (otherwise the integrals diverge).  

The above relations are in the form of a double Laplace transform, namely
\begin{equation}   \label{e9z}
g(s_+,s_-) = \int_{\mathbb{R}^{+\,2}} \D u^+\, \D u^-\, G(u^+,u^-) \, e^{- s_+ u^+ - s_- u^-}\,, 
\end{equation}
where
\begin{equation}   \label{e9a}
 G(u^+,u^-) := -\frac{2\,i}{\pi}\,(u^+ -u^-)\,\tilde\xi(u^+,u^-) \qquad{\rm and}\qquad (k_+-k_-) \, \xi(k_+,k_-) = g(0-i k_+,0-i k_-) \,.
\end{equation}
Invoking now a well known property of the Laplace transform \cite{Vladimirov}, we have that, if 
%\begin{equation}   \label{e9b}
$$ \int_{\mathbb{R}^{+\,2}} \D u^+\, \D u^-\, G(u^+,u^-) \, e^{- a_+ u^+ - a_- u^-} < \infty \,\,\qquad \mbox{for some real} \quad\,a_\pm\,,  $$
%\end{equation}
then $\,g(s_+,s_-)\,$ is analytic in the product of half-planes $\, \realp\left( s_\pm\right) > a_\pm\,$ which, recalling (\ref{e9a}), means that
\begin{equation}   \label{e10}
 (k_+-k_-) \, \xi(k_+,k_-)\quad \mbox{is doubly analytic if} \quad \imp\left( k_+\right) > a_+ \quad{\rm and} \quad \imp\left( k_-\right) > a_-
\end{equation}
Now we need to make a detour to establish some preliminary results that will be helpful in proving the existence of the double Laplace transform $\,g(s_+,s_-)\,$ for $\realp(s_\pm) > 0\,$.

\subsection{Symmetries \label{S1.1} }
Let us now examine the symmetries of the functions $\,\tilde\xi(u^+,u^-)\,$ and $\, \xi(k_+,k_-)\,$ for real values of the variables. 
Using the new variables (\ref{e5a}), the extension (\ref{e4a}) introduced above reads
\begin{equation}   \label{e7}
 \tilde\xi(u^+,u^-) = \tilde\xi(u^-,u^+) \,, 
\end{equation}
therefore $\tilde\xi(u^+,u^-) \,$ is symmetric with respect to the main diagonal. Now the FS-causality condition (\ref{e4}) combined with the extension (\ref{e4a}) implies that $\,\tilde\xi(u^+,u^-)\,$ vanishes outside the first quadrant, $ \,[0, \infty)\times[0,\infty)\, $.

Furthermore, by merely recalling equation (\ref{e6}), we easily find that the symmetry (\ref{e7}) also holds for the Fourier transform
\begin{equation}   \label{e8a}
  \xi(k_+,k_-) =  \xi(k_-,k_+)
\end{equation}
and, since $\,\tilde\xi(u^+,u^-)\,$ must be real for real $u^\pm\,$, the complex conjugate of the relation (\ref{e6}) leads to
\begin{equation}   \label{e8b}
  \xi(-k_+, - k_-) =  \xi^\ast(k^\ast_+,k^\ast_-)
\end{equation}
where the superscript ${\,}^\ast \,$ means ``complex conjugate''.
%In terms of $\,h(k_+,k_-)\,$, for real values of the variables the relations (\ref{e8a}) and (\ref{e8b}) become
%\begin{equation}   \label{e8c}
% h(k_-,k_+) = - h(k_+,k_-) \qquad {\rm and} \qquad  h(- k_+, - k_-) = - h^\ast(k_+,k_-)  
%\end{equation}

\subsection{ The susceptibility function %$\xi(k_+,k_-)\,$ 
at high frequencies \label{S1.2.1}} 
In order to determine the asymptotic behavior of $g(s_+,s_-)$ for large values of $\,s_\pm\,$ we use the following theorem which is proved in Appendix B.

\begin{theorem} \label{t1}
Let $\, g(s_+,s_-)\,$ be the double Laplace integral (\ref{e9z}), assume that $\,G(u^+,u^-)\,$ has continuous partial derivatives $\,G^{(l,j)}(u^+,u^-) = \partial_+^l \partial_-^j G(u^+,u^-)\,$ up to the $N$-th order 
(i. e. $l+j\leq N $) and that there exist $\,a_\pm \in \mathbb{R} \,$ such that 
$\;\int_{\mathbb{R}^{+ 2}} \D u^+ \,\D u^-\, G^{(l,j)}(u^+,u^-) e^{-a_+u^+ - a_- u^-} < \infty\,, \qquad \quad l + j \leq N  \,$,
then
$$ g(\lambda s_+,\lambda s_-) = \sum_{0\leq l+j\leq N-1} \frac{G^{(l,j)}(0^+,0^+)}{\lambda^{l+j+2} s_+^{l+1} s_-^{j+1}} + o(\lambda^{-N-1}) \,, \qquad \qquad \forall \lambda \in \mathbb{R}^+ \,.$$
\end{theorem}

We then say that the asymptotic behavior of $g(s_+,s_-) $ up to order $N-1$ is 
\begin{equation}   \label{e12aa}
 g(s_+,s_-) \sim \sum_{0\leq l+j\leq N-1}  \,\frac{G^{(l,j)}(0^+,0^+)}{s_+^{l+1} \,s_-^{j+1}}
\qquad {\rm when} \qquad s_\pm \rightarrow\infty \,, 
\end{equation}
where $\, G^{(l,j)}(0^+,0^+) \,$ are lateral partial derivatives of $G$.

In order to derive the coefficients of the asymptotic expansion we differentiate (\ref{e9a}) and, on iterating the Leibniz rule, we obtain that 
$$ G^{(0,0)}(0^+,0^+) = 0\,, \qquad \qquad  G^{(1,0)}(0^+,0^+) =- G^{(0,1)}(0^+,0^+) = - \frac{2\,i}{\pi}\,\xi(0^+,0^+) \qquad {\rm and} $$
\begin{equation}   \label{e12ab}
 G^{(l,j)}(0^+,0^+) = -\frac{2 i}{\pi}\,\left[l\,\tilde\xi^{(l-1,j)}(0^+,0^+) - j\,\tilde\xi^{(l,j-1)}(0^+,0^+) \right] 
\end{equation}
Thus the lowest order in the asymptotic expansion (\ref{e12aa}) is 
$$ g(s_+,s_-) \sim   \frac{2\,i \,\tilde\xi(0^+,0^+)}{\pi}\,\frac{s_+ - s_-}{s_+^{2} \,s_-^{2}}  $$
and, using  (\ref{e9a}), we have that
\begin{equation}   \label{e12ac}
 \tilde\xi(k_+,k_-) \sim \frac{2 \,\tilde\xi(0^+,0^+)}{\pi\, k_+^{2} \,k_-^{2}}  \qquad {\rm or} \qquad 
\chi(q,\omega) \sim \frac{2\,\tilde\xi(0^+,0^+)}{\pi (q^2-\omega^2)^2} \, ,
\end{equation}
provided that the partial derivatives $\,G^{(l,j)}(u^+,u^-) = \partial_+^l \partial_-^j G(u^+,u^-)\,$ are continuous for $l,\,j \leq 2\,$ and $u^\pm >0\,$, and $\,(0^+,0)\,$ means the limit for $(t,r) \rightarrow (0,0)\,$, when  $\,|r| < t\,$.

One might require that $ \tilde\xi(u^+,u^-)\,$, which vanishes when either $u^+$ or $u^-$ are negative, does not start abruptly, in other words, the function is continuous at the boundary\footnote{This assumption has been used in \cite{Jackson1}, \S7.10 on the basis that the contrary seems to be counterintuitive; we will see however that it is not so realistic as it might seem.} 
$$ \tilde\xi(u^+,0) = \tilde\xi(0,u^-) = 0 \qquad \mbox{ and therefore} \qquad  \tilde\xi^{(l,0)}(0,0) = \tilde\xi^{(0,j)}(0,0) = 0 \,,$$ which, substituted in (\ref{e12ab}), implies that
\begin{equation}  \label{e11a}
G^{(l,0)}(0,0) = G^{(0,j)}(0,0) = 0 \,, \qquad  G^{(2,1)}(0^+,0^+) = - G^{(1,2)}(0^+,0^+)  = -i\,\frac{4}{\pi c}\,\tilde\xi^{(1,1)}(0^+,0^+)\,. 
\end{equation}
Hence the lowest order in the asymptotic expansion (\ref{e12aa}) is
$$ g(s_+,s_-) \sim -i \frac{4}{\pi}\,\tilde\xi^{(1,1)}(0^+,0^+)\,\frac{s_- - s_+}{s_+^{3} s^3_{-}}$$
and, including (\ref{e9a}), we have that
\begin{equation}   \label{e12a}
 \xi(k_+,k_-) \sim - \frac{4\,\tilde\xi^{(1,1)}(0^+,0^+)}{\pi\,k_+^{3} k_{-}^3 }\,\qquad {\rm or} \qquad \chi(q,\omega) \sim - \frac{4 }{\pi\,(\omega^2- q^2)^{3}}\,\left(\partial_t^2 \tilde\chi - \partial_r^2 \tilde\chi\right)_{(0^+,0)} \,.
\end{equation}

%%%%%%%%%%%%%%%%%%%%%%%%%%%%%%%%%%%%%%%%%%%%%%%%%%%%%%%%%%%%%%%%%%%%%%%%%%%%%%%%%%%%%%%%%%%%%%%
\subsection{The existence of the Laplace transform $g(s_+,s_-)$  \label{S1.2.2}} 
We expect that a constant uniform electric field produces a finite polarization in a non-conductor. If the electric field in equation (\ref{e1}) is constant, we have that the polarization is also a constant,
$$ \mathbf{P} = \frac1\pi\,\mathbf{E}\, \int_0^\infty  \D t\, \int_0^t \D r\, r^2 \tilde\chi(t,r) 
  =  \frac1{\pi c}\,\mathbf{E}\, \int_{\mathbf{R}^+}  \D u^+\, \int_{\mathbf{R}^+} \D u^-\, (u^+-u^-)^2 \,\tilde\xi(u^+,u^-)  \,,  $$ 
where the definition (\ref{e9a}) has been included.
Therefore $	\mathbf{P} $ is finite if, and only if, the integral is finite, that is the function
\begin{equation}   \label{e11c}
  F(u^+,u^-) = \tilde\xi(u^+,u^-)\, (u^+-u^-)^2 = G(u^+,u^-)\, (u^+-u^-) 
\end{equation}
is summable in $\mathbb{R}^{+\, 2}$, which implies that the double Laplace transform   
$$\,f(s_+,s_-) = \int_{\mathbf{R}^{+\,2}}  \D u^+\,  \D u^- \,F(u^+,u^-)\,e^{-(s_+u^+ + s_- u^-)} \,$$ 
is analytic in the product of complex half-planes $\realp(s_\pm) > 0$.  

Proceeding similarly as in section \ref{S1.2.1}, we arrive at the asymptotic expansion 
$$ f(s_+,s_-) \sim \,\frac{4 i}{\pi \,s_+^2 s_-^2}\,\tilde\xi(0^+,0^+) \rightarrow 0\,, \quad \mbox{for large} \quad \,s_+,\,s_- \,.$$
On the other hand the Laplace transform of the relation (\ref{e11c}) implies that
\begin{equation}   \label{e11d}
  f(s_+,s_-) = - \left(\partial_+ - \partial_-\right)\,g(s_+,s_-)\,
\end{equation}
($\partial_\pm$ means the partial derivative with respect to $s_\pm\,$), which on integration yields
\begin{equation}   \label{e11e}
    g(s_+,s_-) = i \, \int_0^\infty \D \sigma\,f(s_+ + \sigma ,s_- - \sigma)  \,,
\end{equation}
where the asymptotic behaviour of $\,f(s_+,s_-)\,$ has been used.
As far as $\,\realp(s_\pm) > 0\,$, the value $\,f(s_+ + i\sigma ,s_- - i \sigma) \, $ is well defined and, due to its asymptotic behaviour, the integral in (\ref{e11e}) converges. Therefore $\, g(s_+,s_-)\,$  exists and is doubly  holomorphic in the region $\,\realp(s_\pm) > 0\,$. 

As a consequence the function $\,(k_+ - k_-)\,\xi(k_+,k_-) = g(0-ik_+,0-ik_-)$ is analytic in the region $\,\imp(k_\pm) > 0\,$, which in terms of the frequence $\omega$ and wave vector $q$ implies that
\begin{equation} \label{e11f}
q\,\chi(q,\omega) \quad \mbox{is doubly holomorphic in} \quad \left|\imp(q)\right| < \imp(\omega) 
\end{equation}
or, equivalently,
\begin{equation} \label{e11g}
\mbox{all singularities of} \quad
q\,\chi(q,\omega) \quad \mbox{are in the region} \quad \imp(\omega) \leq \left|\imp(q)\right|  
\end{equation}

%%%%%%%%%%%%%%%%%%%%%%%%%%%%%%%%%%%%%%%%%%%%%%%%%%%%%%%%%%%%%%%%%%%%%%%%%%%%%%

\subsection{The optical approximation \label{S4a}}
%  on at the closing of Section \ref{S0}, FS-causality is also fulfilled in the optical approximation, when L-causality is required. 
In the optical approximation the susceptibility function depends only on the angular frequency, $\chi(\mathbf{q},\omega) = \chi(\omega)\,$ and its Fourier transform is
\begin{equation}   \label{e11h}
    \tilde\chi(\mathbf{x},t) = (2 \pi)^{3/2}\,\tilde\chi(t)\,\delta(\mathbf{x}) \,.
\end{equation}
When $ \tilde\chi(\mathbf{x},t) $ factorizes like this, L-causality is equivalent to FS-causality; indeed:
\begin{description}
\item[L-causality $\Rightarrow$ FS-causality:] If $t< |\mathbf{x}|\,$, then either $|\mathbf{x}|\neq 0\,$ and $\tilde\chi(\mathbf{x},t) = 0$ due to the $\delta$-function, or $|\mathbf{x}|= 0\,$ and, as $\tilde\chi(t) = 0$ due to to L-causality, then $\,\tilde\chi(\mathbf{x},t)=0\,$.
\item[FS-causality $\Rightarrow$ L-causality:] If $t<0$, then $t< |\mathbf{x}|\,$, for any $\mathbf{x}\,$ and FS-causality implies that $\,\tilde\chi(\mathbf{x},t)=0\,$ for any $\mathbf{x}$ and, including (\ref{0c}), this amounts to  $\tilde\chi(t)= 0$.
\end{description}

We might reason similarly in terms of the Fourier space: 
\begin{itemize}
\item L-causality means that all singularities of the dielectric function $\chi(q,\omega) = \chi(\omega)$ lie in the region $\imp(\omega) \leq 0$, that is $\imp(\omega) \leq |\imp(q)|\,$, for all $q$, which implies FS-causality and,  conversely
\item FS-causality implies that all singularities of  $\chi(q,\omega)$ are in the region $\imp(\omega) \leq |\imp(q)|\,$ and, in particular, all singularities of $\chi(\omega) = \chi(q=0,\omega)$ are in 
$\imp(\omega) \leq 0\,$, which amounts to L-causality.
\end{itemize}

As for the asymtotic behavior, substituting (\ref{e11h}) in (\ref{e9a}), we have that
$$ G(u^+,u^-) = - \frac{4\,i}{\sqrt{2\pi}}\,\tilde\chi\left(u^+ +u^- \right)\,\delta^\prime(u^- - u^+)  \,.$$
As it contains the derivative of a $\delta$-function, it is not a continuous function at the points $\, u^+=u^-\,$ in the quadrant $ \mathbb{R}^+ \times  \mathbb{R}^+\,$, the hypothesis of Theorem \ref{t1} are not met and therefore $\chi(\omega)$ may not present the asymptotic behavior (\ref{e12a}). 

%%%%%%%%%%%%%%%%%%%%%%%%%%%%%%%%%%%%%%%%%%%%%%%%%%%%%%%%%%%%%

\section{A generalization of  Kramers-Kr\"onig relations \label{S4}  }
In the optical approximation Kramers-Kr\"onig relations connect the real and imaginary parts of electric susceptibility $\chi(\omega)$ in such a way that knowledge of one for all real values of $\omega$ determines the other: they are Hilbert transforms of each other \cite{Titchmarsh}. These relations follow from: (i) the fact that $\chi(\omega)$ is holomorphic in the upper complex half-plane $\imp(\omega) > 0\,$ combined with (ii) a suitable asymptotic behavior for large $|\omega|\,$. 
If there is spatial dispersion, FS-causality implies that $\,(k_+-k_-)\,\xi(k_+,k_-) = 2 q\,\chi(q,\omega)\,$ is holomorphic in the product of half-planes $\,\imp(k_\pm) > 0\,$. 

Given two real values $\,k_\pm\,$, consider the closed path $\mathcal{C}_{R,\delta}$ in the upper complex half-plane, consisting of the upper half-circle centered at $k_++i \varepsilon\,$ with radius $R$ and the line $\,k_+ + i \varepsilon + t\,$, $\,|t| < R\,$. Since $\,(k_+-k_-)\,\xi(k_+,k_-)\,$ is holomorphic on and inside $\mathcal{C}_{R,\delta}$ and $k_+$ is outside this region, Cauchy-Goursat theorem \cite{Detman} implies that
\begin{equation} \label{kk0}
  \oint_{\mathcal{C}_{R,\delta}} \D k^\prime\, \frac{\xi(k^\prime, k_-)}{k^\prime - k_+}\,(k^\prime- k_-) = 0 
%\,, \qquad \imp(k_+) \geq 0 \,, \quad  \imp(k_-) > 0
\,.
\end{equation}
Then, provided that $\;\lim_{|k^\prime|\rightarrow\infty}(k^\prime-k^{\prime\prime})\,\xi(k^\prime,k^{\prime\prime}) =: H(k^{\prime\prime})\,, \;\,  0 < {\rm arg}\,k^\prime < \pi \,$, exists, we have that in the limit $\,R\rightarrow\infty\,, \;\, \delta \rightarrow 0\,$, the above integral leads to
\begin{equation}  \label{Circ2}  
(k_+-k_-)\,\int_{\mathbb{R}+ i 0}\frac{ \D k^\prime}{k^\prime - k_+} \, \xi(k^\prime, k_-) + \int_{\mathbb{R}} \D k^\prime \, \xi(k^\prime, k_-) + i \pi\, H(k_-) = 0\,, 
\end{equation}
where $\mathbb{R}+ i 0$ means the limit of $\mathbb{R}+ i \varepsilon\,$ for $\,\varepsilon\rightarrow 0^+\,$.
Taking $k_+=k_-$, this yields 
$$ \int_{\mathbb{R}} \D k^\prime \, \xi(k^\prime, k_-) + i \pi\, H(k_-) = 0 \, $$
and therefore
$$ \int_{\mathbb{R}+ i 0}\frac{ \D k^\prime}{k^\prime - k_+} \, \xi(k^\prime, k_-) = 0 \,, \qquad {\rm if}\quad  k_+ \neq k_- \,.$$
Using the Sochozki formula \cite{Vladimirov}, this equation readily leads to
\begin{equation}   \label{e16z}
 \xi(k_+,k_-) = \frac1{i \pi}\,\mathcal{P}\int_{\mathbb{R}} \frac{ \D \rho}{\rho} \,\xi(k_+ +\rho, k_-) = \frac1{i \pi}\,\mathcal{P}\int_{\mathbb{R}} \frac{ \D \rho}{\rho-k_+} \,\xi(\rho, k_-)  \,, 
%\qquad  \imp(k_+) \geq 0 \,, \quad  \imp(k_-) > 0  \,,
\end{equation} 
which for future use we can equivalently write as
\begin{equation}   \label{kk1}
 \xi(k_+,k_-) = \frac{i}{2 \pi}\,\int_{\mathbb{R}} \frac{ \D \rho}{\rho + i 0} \,\xi(k_+ -\rho, k_-) = \frac{i}{2 \pi}\,\int_{\mathbb{R}} \frac{ \D \rho}{k_+ - \rho + i 0} \,\xi(\rho, k_-) \,, 
%\qquad  \imp(k_+) \geq 0 \,, \quad  \imp(k_-) > 0  \,,
\end{equation}

We can proceed similarly with the variable $k_-$ and obtain, instead of (\ref{e16z}),
\begin{equation*}  % \label{e16a} 
 \xi(k_+,k_-) = \frac1{i \pi}\,\mathcal{P}\int_{\mathbb{R}} \frac{ \D \rho}{\rho-k_-} \,\xi(k_+,\rho) \,, 
%\qquad   \imp( k_-) \geq 0 \,, \quad  \imp(k_+) > 0  \,,   
\end{equation*} 
which does not add anything new because it is equivalent to (\ref{e16z}), by the symmetry relation (\ref{e8a}) and, applying it 
%(\ref{e16a}) 
after (\ref{e16z}), we also obtain 
\begin{equation}   \label{kk2}
 \xi(k_+,k_-) = -\frac1{\pi^2}\,\mathcal{P}\int_{\mathbb{R}} \frac{ \D \rho}{\rho-k_+} \,\mathcal{P}\int_{\mathbb{R}} \frac{ \D \tau}{\tau - k_-} \,\xi(\rho, \tau) \,.
%\qquad  \imp(k_+) \geq 0 \,, \quad  \imp(k_-) > 0  \,,
\end{equation} 

Similarly, starting form (\ref{kk1}) and using the symmetric relation for the $\,k_-\,$ in the rhs, we have the equivalent relation 
\begin{equation}   \label{kk3}
 \xi(k_+,k_-) = -\frac1{4 \pi^2}\,\int_{\mathbb{R}} \frac{ \D \rho}{k_+ - \rho + i 0} \,\int_{\mathbb{R}} \frac{ \D \tau}{k_- - \tau + i 0} \,
\xi(\rho,\tau) \,.
% \qquad  \imp(k_+) \geq 0 \,, \quad  \imp(k_-) > 0  \,,
\end{equation}

So far we have derived four integral relations on $\, \xi(k_+,k_-) \,$, namely (\ref{e16z}-\ref{kk3}), with different relevance.
Just like Kramers-Kr\"onig relations, equation (\ref{e16z}) allows to obtain the real part of $\xi$ in terms of its imaginary part, and conversely. On the contrary, the real and imaginary parts of (\ref{kk2}) respectively act as constraints on $\realp(\xi)$ and $\imp(\xi)$, separately. Thus (\ref{e16z}) implies (\ref{kk2}) but the converse is not true.

In turn, each of the relations (\ref{e16z}), (\ref{kk1}) and (\ref{kk3}) mixes the real and imaginary parts of $\xi\,$. As commented before, the relations (\ref{e16z}) and (\ref{kk1}) are equivalent, and (\ref{kk3}) is equivalent too thanks to the fact that
\begin{equation}   \label{e16q}
 \int_{\mathbb{R}} \frac{ \D \lambda}{(k_+ - \lambda + i 0)(\lambda - \rho + i 0)} = - \frac{2 \pi\, i}{k_+ - \rho + i 0} \,,
\end{equation}
which implies that  (\ref{kk1}) holds if, and only if,  (\ref{kk3}) does.
%%%%%%%%%%%%%%%%%%%%%%%%%%%%%%%%%%%%%%%%%%%%%%%%%%%%%%%%%%%%%%%%%%%%%%%%%%%%%%%%

For the sake of comparison with other extensions of Kramers-Kr\"onig relations, it is useful to write equation (\ref{kk3}) in terms of $\chi(q,\omega) \,$ as
\begin{equation}   \label{kk4}
 \chi(q,\omega) = -\frac1{4 \pi^2}\,\int_{\mathbb{R}} \frac{ \D \rho}{\rho + i 0} \,\int_{\mathbb{R}} \frac{ \D \tau}{\tau + i 0} \,
\chi\left(q +  \frac{\tau-\rho}2,\,\omega - \frac{\rho+\tau}2 \right) \,
\end{equation}
or, equivalently, 
\begin{equation}   \label{kk4a}
 \chi(q,\omega) = -\frac1{2 \pi^2}\,\int_{\mathbb{R}^2} \D q^\prime \D\omega^\prime\,\frac{\chi\left(q^\prime,\,\omega^\prime\right)}{(\omega -\omega^\prime + q - q^\prime + i 0) \,(\omega -\omega^\prime - q + q^\prime + i 0)} \,.
\end{equation}

In terms of the variables $\,(q,\omega)\,$, relations (\ref{e16z}) and (\ref{kk1}) respectively read
\begin{equation}   \label{e16b}
\chi(q,\omega) = \frac1{i \pi}\,\mathcal{P}\int_{\mathbb{R}} \frac{\D \zeta}{\zeta}\,\chi\left(q +\zeta,\omega + \zeta \right)  = 
\frac{i}{2 \pi}\,\int_{\mathbb{R}} \frac{\D \zeta}{\zeta+i 0}\,\chi\left(q -\zeta,\omega - \zeta \right)\,, 
\end{equation}
which, including the $q$-parity of $\chi(q,\omega)\,$, implies that
$$ \chi(q,\omega) = \frac1{i \pi}\,\mathcal{P}\int_{\mathbb{R}} \frac{\D \zeta}{\zeta}\,\chi\left(q -\zeta,\omega + \zeta \right)   $$
and, using it in the right hand side of (\ref{e16b}), it yields
\begin{equation}   \label{e16p}
\chi(q,\omega) = -\frac1{\pi^2}\,\mathcal{P}\int_{\mathbb{R}} \frac{\D \zeta}{\zeta}\,\mathcal{P}\int_{\mathbb{R}} \frac{\D \rho}{\rho}\,\chi\left(q +\zeta-\rho,\omega + \zeta+\rho \right)   \, 
\end{equation}
that is equivalent to (\ref{kk2}), as it easily follows from the relation between $\chi$ and $\xi\,$.

Parallely to what has been said concerning the equivalence between (\ref{kk1}) and (\ref{kk3}), it results that (\ref{e16b}) and (\ref{kk4a})
are equivalent too.

\paragraph{Hilbert transform:} 
If we now write $\,\chi = \chi_1 + i \,\chi_2\,$ the relation (\ref{e16b}) can be splitted in its real and imaginary parts
\begin{eqnarray}   \label{e16y}
\chi_1(q,\omega) &=& \frac1{\pi}\,\mathcal{P}\int_{\mathbb{R}} \frac{\D \zeta}{\zeta}\,\chi_2\left(q + \zeta, \omega + \zeta  \right) \quad\qquad {\rm and}  \\[1ex]  \label{e16ya} 
\chi_2(q,\omega) &=& - \frac1{\pi}\,\mathcal{P}\int_{\mathbb{R}}\frac{\D \zeta}{\zeta}\,\chi_1\left(q + \zeta, \omega + \zeta  \right)\,, 
\end{eqnarray}
hence the real part is determined by the imaginary part and conversely. 

Apparently (\ref{e16y}) does not look like a Hilbert transform  \cite{Titchmarsh} and one might infer that applying (\ref{e16ya}) after (\ref{e16y}) would result in an integral constraint on $\chi_1\,$. However, introducing the variables
$$ \lambda = q + \omega + 2 \zeta  \,, \qquad \omega \pm q \qquad \mbox{and the function} \qquad  \xi\left(\omega+q,\omega-q\right) = \chi(q,\omega) \,, $$
equation (\ref{e16y}-\ref{e16ya}) can be written as
\begin{eqnarray*}
\xi_1\left(\omega +q,\omega -q\right) &=& \frac1{\pi}\,\mathcal{P}\int_{\mathbb{R}} \frac{\D \lambda}{\lambda - \left(\omega + q \right)}\,\,\xi_2\left(\lambda, \omega -q\right)  \\[2ex]   
\xi_2\left(\omega +q,\omega -q\right) &=& - \frac1{\pi}\,\mathcal{P}\int_{\mathbb{R}} \frac{\D \lambda}{\lambda - \left(\omega + q \right)}\,\,\xi_1\left(\lambda, \omega -q \right) \,.
\end{eqnarray*}
In other words, $\,\displaystyle{\xi_1\left(\omega +q,\omega -q\right)}\,$ is the Hilbert transform of $\,\displaystyle{\xi_2\left(\omega +q,\omega -q\right)}\,$. Although this fact is concealed in the expressions  (\ref{e16y}-\ref{e16ya}), the use of the right variables has made it apparent. As a consequence, both relations are the inverse of each other, similarly to what happens with the standard Kramers-Kr\"onig relations, because the Hilbert transform is anti-involutive and the supposed restriction is trivial.

Furthermore, introducing the variable, $\, \zeta = \tau - \omega\,$, equation (\ref{e16b}) becomes
\begin{equation}   \label{e16m}
 \chi(q,\omega) = \frac1{i \pi}\,\mathcal{P}\int_{\mathbb{R}} \frac{\D \tau}{\tau - \omega}\,\chi\left(q + \tau - \omega, \tau  \right)  \,,
\end{equation}
which looks like the relation obtained by Leontovitch \cite{Leontovich1961} for the conductivity $\,\sigma(q,\omega)\,$ as a relativistic generalisation of Kramers-Kr\"onig relations. We will get back to this point in Section \ref{S6}.

%%%%%%%%%%%%%%%%%%%%%%%%%%%%%%%%%%%%%%%%%%%%%%%%%%%%%%%%%%%% 
\subsection{The standard Kramers-Kr\"onig relations \label{SS3.1}}
For all real values of $q$, the susceptibility function $\chi(q,\omega)\,$ fulfills the standard K-K relations. Indeed, as $\imp(q)=0\,$, the property (\ref{e11f}) implies that $\chi(q,\omega)\,$ is holomorphic in the upper half-plane $\,\imp(\omega) > 0\,$ and, provided that the necessary asymptotic conditions are granted, the standard Kramers-Kr\"onig relations follow \cite{Jackson1}:
\begin{equation}  \label{e17}
 \chi(q,\omega) = \frac1{i\,\pi}\,\mathcal{P}\int_{-\infty}^\infty \frac{\D \omega^\prime}{\omega^\prime-\omega}\,\chi(q,\omega^\prime) \,. 
\end{equation}
Notice that this relation is not a particular case of the generalised relation (\ref{e16b}), which has been derived differently, 
namely by integrating over a path in the complex plane $\,\displaystyle{k_+ =\omega + q}\,$ for a fixed $\,\displaystyle{k_- =\omega- q}\,$. Instead, the standard K-K relation is derived by integrating over a path on the $\omega$-plane with fixed real $q\,$. The variable that remains fixed is different in each case.

It is worth saying here that the ref. \cite{Rozanov} claims that, in a certain context, Kramers-Kr\"onig relations are [approximately] applicable in its  conventional form because some ``relativistic factor'' is smaller that $10^{-4}$. In fact, we have just shown here that these relations hold exactly without any relativistic correction.

%%%%%%%%%%%%%%%%%%%%%%%%%%%%%%%%%%%%%%%%%%%%%%%%%%%%%%%%%%%%%%%%%%%%%%%%%%
\subsection{The significance and usefulness of causality conditions}
So far we have derived some necessary conditions to be fulfilled by $\chi(q,\omega)\,$ to be FS-causal. 
\begin{list}
{{\bf (\Alph{llista})}}{\usecounter{llista}} 
\item For complex $q$ and $\omega$, $\chi(q,\omega)\,$ is doubly holomorphic in the region $\,\imp\omega > |\imp q|\,$. 
\item For $\,q,\,\omega \in \mathbb{R}\,$, the integral relations (\ref{e16b}) and (\ref{e16p}) are fulfilled.
\item  For real $\,q,\,\omega\,$, the integral relation (\ref{kk4a}) is met.
\end{list}
(Their counterparts in terms of $\xi(k_+,k_-)\,$ are specially useful because they simplify the algebraic handling.)
These conditions keep a certain hierarchy: the first one implies the other two, whereas {\bf (B)} and {\bf (C)} are equivalent. The interest of {\bf (B)} lies in the fact that (\ref{e16b}) is a sort of disguised Hilbert transform which determines $\imp \chi\,$ from $\realp \chi\,$ and conversely ---pretty much as the conventional Kramers-Kr\"onig relations---, whereas (\ref{e16p}) is a double integral constraint on both $\realp \chi\,$ and $\imp \chi\,$. On the other hand the form {\bf (C)} will be convenient when we compare our results with previous ones in Section \ref{S6}.   

Conditions {\bf (B)} and {\bf (C)} are necessary, but not sufficient, for FS-causality. Indeed, if $\,\chi\,$ behaves appropriately in the asymptotic region of large $\,|\omega\pm q| \,$, then {\bf (A)} implies {\bf (B)} but the converse is not true. Condition (\ref{e11g}) might fail and instead equality (\ref{e16b}) may hold, e. g. if $(k_+-k_-)\,\xi (k_+,k_-)$ had a pole of second order for some point in $\imp(k_+) > 0\,$, which would not contribute the integral (\ref{kk0}). 

The practical use and applicability of each of these FS-causality conditions depend on how deeply we know the susceptibility function.
\begin{enumerate}
\item If we know $\chi(q,\omega)$ in closed form and its singular points can be found by algebra, as it happens for susceptibility functions derived from a microscopic model (see Section \ref{S5}), the most workable is to check whether these singular points lie in the region (\ref{e11g}). This provides a complete test for FS-causality.
\item If $\chi(q,\omega)$ is given in closed form but the algebraic determination of its singular points is not feasible (e.g. \cite{Mermin1970}), we can check (\ref{e16b}) and (\ref{e16p}) by numerical methods. This is only a numerical test of a necessary, but not sufficient, condition for FS-causality.
\item If $\chi(q,\omega)$ is only known numerically for real values of $q$ and $\omega$, we have to resign ourselves to checking (\ref{e16b}) and (\ref{e16p}) by numerical methods.  
\end{enumerate}

%%%%%%%%%%%%%%%%%%%%%%%%%%%%%%%%%%%%%%%%%%%%%%%%%%%%%%%%%%%%%%%%%%%%%%%
\section{Application to some dielectric functions \label{S5}} 
Next we examine some dielectric functions that frequently arise in the literature to check whether they satisfy the FS-causality conditions (\ref{e11f}). We concentrate mostly on the dielectric functions derived in \cite{Lindhard1954}.

\subsection{Degenerate electron gas. Semiclassical model  \label{S5.1} }
The first model presented by Lindhard consists of a Fermi electron gas at zero temperature and the {\em longitudinal} and {\em transverse susceptibility functions}\footnote{See Appendix A for the definition.} are respectively ---equations (2.4) and (2.3) in ref. \cite{Lindhard1954}---
\begin{equation}  \label{ap1}
\chi^L (q,\omega)  = \frac{3\omega_{\rm p}^2 m }{4 \pi v_0 p_0 \,q^2} \,\left[ 1 + \frac{\omega+i/\tau}{2 v_0 q } \,\log\psi \right]  
\end{equation}
and 
\begin{equation}   \label{ap1a}
\chi^T (q,\omega) =  \frac{3\omega_{\rm p}^2 m}{16 \pi p_0 \:\omega \:q} \,\left( -2 \,\frac{\omega+i/\tau}{v_0 q} + \left[1-\left(\frac{\omega+i/\tau}{v_0 q }\right)^2\right] \,\log\psi \right) 
\end{equation} 
where
$$ \psi = \frac{\omega+i/\tau - v_0 q}{\omega+i/\tau + v_0 q} \,, \qquad \qquad \omega_{\rm p} =\sqrt{\displaystyle{ \frac{4\pi e^2 n}{m}}} $$
is the plasma frequency, $n$ is the electron density, $\,\displaystyle{ p_0 = h\,\left(\frac{3 \,n}{4\pi}\right)^{1/3} }\,$ is the Fermi momentum and $\,v_0\,$ is the corresponding speed.
They apply to both the relativistic and the non-relativistic case, which differ only in the choice of the velocity-momentum relation. In the non-relativistic model $\,\displaystyle{ v_0 = \frac{p_0}{m} }\,$  is unbounded, whereas in the non-relativistic case  
$\displaystyle{ v_0 = \frac{p_0}{\sqrt{m^2 +p_0^2}} < 1}\,$.

The singularities of the electrical susceptibilities are:
\begin{description}
\item[(i)] branch points at $\omega +i/\tau\pm v_0 q = 0\,$, which means that 
$$ \omega_2 = -\frac1\tau \pm v_0\,q_2\,, \qquad {\rm with} \qquad
\omega_2=\imp(\omega)\,, \qquad q_2 = \imp(q)\,.$$
\begin{figure}[htb]  \label{F2}
\begin{center}
\includegraphics[width=7cm]{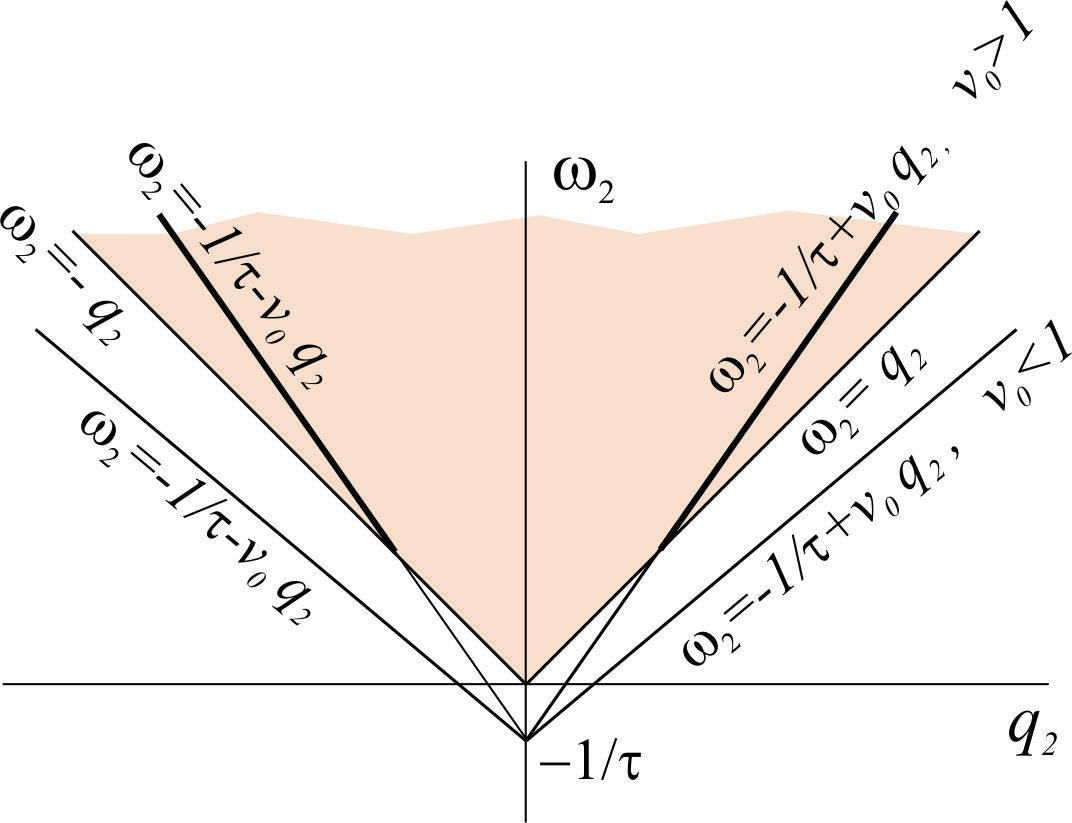}
\end{center}
\caption{The branch points are arranged on the lines $ \omega_2 = -1/\tau \pm v_0\,q_2 $ which intersect the forbidden region (\ref{e11f}), i. e. $\omega_2 > |q_2|$, if and only if $v_0>1$. }
\end{figure}

If $p_0$, $m$ and $v_0$ are connected by the relativistic relation, then  $v_0 < 1\,$ and a simple inspection of Figure 1 reveals that the branch points lie inside the permitted region (\ref{e11g}). In turn, if we choose the non-relativistic relation, then $v_0$ is not bounded and, for high enough electron densities (large $p_0$), $v_0 > 1 $ and the region $\imp(\omega)>|\imp(q)|$ contains an infinite number of branch points ---as it is illustrated in Figure 1--- and FS-causality is violated. 
Note, however, that if the Fermi velocity is $v_0 < 1\,$, even the non-relativistic model has no such branch points, even though relativistic causality has nowhere been imposed at the microscopic level.
\item[(ii)] Both functions seem to have a pole at $q=0$ but this is a false singularity in both cases. Indeed, as $\omega + i/\tau\neq 0\,$, we can substitute the Taylor expansion of the logarithm around $q=0$ in equations (\ref{ap1}-\ref{ap1a}) to obtain neat Taylor series 
$$ \chi^L (q,\omega) = - \frac{\omega_{\rm p}^2 m v_0}{4 \pi p_0 \,\left(\omega+i/\tau\right)^2}+ O(q^2) \qquad {\rm and} \qquad 
\chi^T (q,\omega) = - \frac{\omega_{\rm p}^2 m v_0}{4 \pi p_0 \,\omega \left(\omega+i/\tau\right)}+ O(q^2)\,,$$
and finally
\item[(iii)] the transverse function presents a pole at $\omega = 0\,$ which lies inside the allowed region (\ref{e11g}).
\end{description} 
If  $q=0$, both susceptibilities present a pole at $\omega= -i/\tau\,$ but, if $\tau\neq\infty\,$,
$\;\displaystyle{ \imp(\omega) = - \frac1\tau \leq 0 =|\imp(q)| \,}$,
which lies inside the allowed region (\ref{e11g}).

%%%%%%%%%%%%%%%%%%%%%%%%%%%%%%%%%%%%%%%%%%%%%%%%
The right asymptotic behavior of $q\,\chi(q,\omega)\,$ for large $k_\pm\,$ is crucial to derive the integral relation (\ref{e16b}) from the fact that $q\,\chi(q,\omega)$ has no singularities in the region (\ref{e11f}), as it happens when $v_0 < 1$. Substituting the relations (\ref{e5b}) into equations (\ref{ap1}-\ref{ap1a}) in the non-relativistic regime ($v_0 \ll 1$), this  leads to the asymptotic expansion
$\,  (k_+ - k_-)\,\xi^A(k_+,k_-) = q\,\chi^A(q,\omega) = O(k_+^{-1}) \,, \quad A=L,\,T\,$, which is sufficient to derive the integral expression (\ref{Circ2}) with $H(k_-)=0\,$.

%In the present case a straight evaluation, which involves Spence dilogarithm functions, shows that both $\chi^L$ and $\chi^T$ satisfy the integral relation (\ref{Circ2}).

It is worth noticing that the functions (\ref{ap1}-\ref{ap1a}) do not meet the asymptotic behavior (\ref{e12ac}). This is because the hypothesis of Theorem \ref{t1} is not met. Indeed, the latter includes the vanishing of $G(u^+,u^-)$ when any of the two variables is negative, i. e. FS-causality, and its continuity in this domain. 
Restricting to the FS-causal case $\tau=\infty$ and $\,v_0 < 1\,$, the inversion of the Fourier transform (\ref{e5}) for the longitudinal function is a short calculation  that yields
$$ \tilde\chi^L(r,t) = \frac{3 \omega_p^2 m}{4 p_0 v^2_0 t^2}\,\Theta(v_0 t-|r|) \,.$$
This mets the FS-causality condition if, and only if, the Fermi velocity does not exceed the speed of light in vacuum. If $v_0 < 1\,$, $ \tilde\chi^L(r,t) $ is not a continuous function in the region $\,t \geq |r|\,$, i. e. $u^\pm \geq 0\,$ and, if $v_0=1$ the second term is a continuous function that starts abruptly at the boundary $t = |r|\,$.

\subsection{Non-relativistic quantum degenerate electron gas (random phase approximation)  \label{S5.3} }
Lindhard \cite{Lindhard1954} also derives both the longitudinal and transverse dielectric functions for the non-relativistic quantum model of a Fermi electron gas at zero temperature. The susceptibilities are
\begin{equation}  \label{ap3} 
 \chi^L(q,\omega) =    \frac{3 \omega_{\rm p}^2 }{8 \pi\, q^2 v_0^2} \, f^L(z,u)   \qquad {\rm and} \qquad 
\chi^T(q,\omega) =  - \frac{3 \omega_{\rm p}^2 }{32 \pi\,  \omega^2} \, f^T(z,u)
\end{equation}
with
\begin{equation}  \label{ap3b} 
 f^L(z,u) = 1 + \frac{1 - (z-u)^2}{4z}\, \log \frac{z - u +1}{z - u -1} + \frac{1 - (z+u)^2}{4z}\, \log \frac{z + u +1}{z + u -1}  
\end{equation}
and 
\begin{equation}  \label{ap3c} 
f^T(z,u) = 1 + 3 u^2 + z^2 - \frac{\left[1 - (z-u)^2\right]^2}{4z}\, \log \frac{z - u +1}{z - u -1} - \frac{\left[1 - (z+u)^2\right]^2}{4z}\, \log \frac{z + u +1}{z + u -1}  
\end{equation}
where the dimensionless variables
\begin{equation}  \label{ap3a} 
 z = \frac{q}{2 q_0}\,, \qquad u = \frac{\omega^\prime}{ q \,v_0} \,, \qquad \omega^\prime = \omega + i \frac{\gamma}{\hbar} \,, 
\end{equation}
have been used and $v_0$ and $\hbar q_0\,$ are the values of the velocity and momentum at the Fermi surface.

As for the singularities of $ \chi(q,\omega) \,$ we find that:
\begin{description}
\item[\underline{$q=0$},] or $z=0\,$, is only an apparent singularity of $\chi^L\,$. Indeed, puting $u = y/z\,$, with $\,\displaystyle{ y = \frac{\omega^\prime}{2\, q_0 \,v_0} }\,$, writing the logarithm terms as
$$ \log \left(1 \mp \displaystyle{\frac{z^2 + z}{y}}\right) - \log \left(1 \mp \displaystyle{\frac{z^2 - z}{y}}\right) $$
and taking the Taylor expansion $\log(1+x) = x - x^2/2 + O(x^3) \,$, we easily arrive at
$\,f^L(z,u) = - {z^2}/{y^2} + O(z^3)\, $
and, including (\ref{ap3}), we see that $\,\displaystyle{\chi^L(q,\omega) = -\frac{3\omega_{\rm p}^2}{8\pi \omega^{\prime\,2}} +O(q)}\,$, which has no singularity at $\,q=0\,$.

\item[\underline{$\omega=0$},] is a pole of $\chi^T(q,\omega) \,$, but it is outside the region $\,\imp(\omega) > |\imp(q)|\,$.

\item[Branch points] at  $\,z + \sigma\,u + \lambda = 0\,$, with $\,\sigma,\lambda = \pm 1\,$ which, including (\ref{ap3a}), amounts to
\begin{equation}  \label{ap3bb}
 \omega + i\,\frac{\gamma}\hbar  = \frac{\hbar}{2\sigma m }\,\left( q + \lambda q_0\right)^2 - \frac{\hbar q_0^2}{2\sigma m } \,. 
\end{equation}
To decide whether these branch points do lie in the forbidden region (\ref{e11f}), $\,\omega_2 > |q_2|\,$, we only need to examine the imaginary parts of $\, q = q_1 + i \,q_2$ and $\,\omega = \omega_1 + i \,\omega_2\,$ and check if  
\begin{equation}  \label{ap3bbb}
  -\sigma s_1 s_2 \,\frac{\hbar}{m}\,|q_2|\,\left(|q_1|+\lambda s_1 q_0 \right) = \omega_ 2 + \frac{\gamma}\hbar >|q_2|  + \frac{\gamma}\hbar \,,
\end{equation}
where $\,s_j= {\rm sign}\,q_j\,$.

This inequality implies that $|q_2|\neq 0\,$ and that $\,\lambda s_1 q_0 +|q_1| > 0\,$. Indeed, otherwise it should be $ \lambda s_1 = -1\,$ and $ q_0 > |q_1|\,$, then (\ref{ap3bbb}) would imply that $-\sigma s_1 s_2 = -1 \,$ and $\,\hbar q_0 > mc\,$, which lies outside the non-relativistic regime.
Then, combining  $\,\lambda s_1 q_0 +|q_1| > 0\,$ with (\ref{ap3bbb}), we arrive at 
$$ -\sigma s_1 s_2 = +1 \qquad {\rm and} \qquad 
|q_2|\,\left(|q_1| -\frac{m}\hbar + \lambda s_1 q_0 \right) > \frac{\gamma m}{\hbar^2} \,. $$
This corresponds to the outer region of an equilateral hyperbola in the quadrant $|q_2|>0\,$, $|q_1| >\frac{m }\hbar - \lambda s_1 q_0\,$. Any $q$ fulfilling these conditions and the corresponding $\omega $ defined by (\ref{ap3bb}) yield a branch point in the forbidden region (\ref{e11f}), therefore Lindhard dielectric functions (\ref{ap3}) violate the FS-causality condition, regardless of the magnitude of the Fermi momentum.
\end{description}
Notice that, as commented in Section \ref{SS3.1}, the latter is consistent with the fact that Lindhard dielectric functions fulfill the Kramers-Kr\"onig relations for real $\,\omega\,$ and $\, q\,$, because in this case $\,q_2= 0\,$ and $\omega_2 < 0\,$, which falls outside the domain (\ref{e11f}).

%%%%%%%%%%%%%%%%%%%%%%%%%%%%%%%%%%%%%%%%%%%%%%%%%%%%%%%%%%%%%%%%%%%%%%%%%%%%%%%%%%%%%%%%%%%%%%%%%%%%%
\subsection{Valence electrons in semiconductors and insulators \label{S5.4} }
Levine and Louie \cite{Levine1982} derived a model dielectric function by adding a lowest excitation frequency (or ``gap'') into the Lindhard  dielectric function $\,\varepsilon^L(q,\omega)\,$. They first modified the imaginary part of the function and used the standard Kramers-Kr\"onig relations to derive the real part and the result is an expression for $\,\varepsilon^L(q,\omega)\,$ for real values of $q$ and 
$\omega\,$. A closed expression for complex valued $q$ and $\omega$ can be obtained by replacing $\omega$ with $\,\sqrt{\omega^2 - \omega_g^2}\,$ in the expressions (\ref{ap3}-\ref{ap3a}). The resulting dielectric function presents the same false pole at $q=0$, two extra branch points at $\pm \omega_g \,$ which is real ($\imp \omega = 0\,$) and other branch points at
$$  \left(q + \lambda q_0\right)^2 -q_0^2 = -\frac{2\sigma m }{\hbar}\,\sqrt{\omega^2 - \omega_g^2} \,,\qquad\quad \lambda,\, \sigma=\pm 1$$
For real values of $q$, the right hand side must be real, which implies that $\omega$ is real as well. 
For $ q = q_1 + i q_2\,, \quad q_2\neq 0\,$, examining the imaginary part of the square of the above equation we easily obtain
$$ \frac{\omega_2}{q_2} = \frac{\hbar^2 (q_1+\lambda q_0)}{m^2 \omega_1}\,\left[(q_1+\lambda q_0)^2 - q_0^2 - q_2^2 \right] \,.$$
For large $q_1\,$ the right hand side can be larger than $1$ and $\omega_2 > q_2\,$, which produces a  branch point in the forbidden region  (\ref{e11f}). Therefore the dielectric function of Levine and Louie \cite{Levine1982} violates FS-causality. 
%%%%%%%%%%%%%%%%%%%%%%%%%%%%%%%%%%%%%%%%%%%%%%%%%%%%%%%%%%%%%%%%%%%%%%%%%%%%%%%%%%%%%%%%%%%%%%%%%%%%%%%%%%

\subsection{Relativistic quantum degenerate electron gas  \label{S5.5} }
The dielectric functions $\varepsilon^L(q,\omega)$ and $\varepsilon^T(q,\omega)$ of an electron gas within a quantum electrodynamics framework (for real positive values of $q$ and $\omega$) were derived by Jancovici \cite{Jancovici1962}. He gave both the real and imaginary parts, respectively equations (A.1) and (A.1') in  \cite{Jancovici1962}, for the longitudinal dielectric function, and (A.4) and (A.4') for the transverse function.

The analytic extension of the longitudinal dielectric function to complex $q$ and $\omega$ is (in natural units $\,\hbar=c=1 $)
$$  \varepsilon^L  = \varepsilon^\prime_0 + \frac{2 e^2 }{3\pi}\left(  \frac{4 q_0^2 \beta_0^{-1}}{q^2} - \log\frac{1+\beta_0^{-1}}{\mu_0}\right) +  
  A\,\log \frac{Q_+}{Q_-}  +  B\,\log \frac{N_+}{N_-}  +  C\,\log \frac{P_+}{P_-} \,,   $$
where $q_0$ and  $ \beta_0 $ are the Fermi momentum and velocity, 
$\, \beta_0^{-1}= \sqrt{1 +\mu_0^2}  \,$,  $\;\mu_0 = m/q_0= \gamma_0^{-1}\,\beta_0^{-1}\,$, and
\begin{eqnarray*}
A  & = & \frac{e^2 (q^2 -\omega^2-2m^2)}{6 \pi (q^2-\omega^2)} \sqrt{ \frac{q^2 -\omega^2+4m^2 }{q^2-\omega^2}} \\[2ex]
Q_\lambda  & = &  \left[(q^2-\omega^2) q_0 \beta_0^{-1} + \lambda q_0\sqrt{(q^2-\omega^2)(q^2-\omega^2+4m^2)}  \right]^2  - 4 m^4 \omega^2 \,, \qquad \lambda=\pm 1\\[2ex]
B  & = &  \frac{e^2\, \omega}{\pi q^3} \,\left(\frac{q^2}{4} - \frac{\omega^2}{12} - q_0^2 \beta_0^{-2}  \right) \,,  %\\[2ex]
\qquad \qquad
N_\lambda   = 4 q_0^2 (q +\lambda \omega \beta_0^{-2}) - (q^2 \omega^2)^2 \\[2ex]
C  & = &  \frac{e^2\, q_0 \beta_0^{-1}}{\pi q^3} \,\left[\frac23 q_0^2 \beta_0^{-2} - \frac12 (q^2-\omega^2) \right] \,,    
\quad  P_\lambda = \left[2 q q_0 + \lambda (q^2 -\omega^2)\right]^2 - 4\omega^2 q_0^2 \beta_0^{-2}   \\[2ex]
\varepsilon^L_0   & = &  1 + \frac{e^2}{3\pi}\,\left(\frac53 -\frac{4 q_0^2\mu_0^2}{q^2-\omega^2} \right) - 4 \,A\,\ln \left(\sqrt{\frac{q^2 -\omega^2}{4m^2}} + \sqrt{\frac{q^2-\omega^2+4m^2 }{4m^2}} \right) 
\end{eqnarray*}   
For the sake of convenience, we translate these expressions into the dimensionless variables 
\begin{equation}    \label{ap3aa}
   z=\frac{q}{2 q_0} \,, \qquad  u^\prime=\frac{\omega}{2 \, q_0} \,\qquad {\rm and } \qquad   \zeta = \frac{\sqrt{z^2 - u^{\prime 2}}}{\mu_0} \,,
\end{equation}
and, after some manipulation, we obtain
\begin{equation}  \label{b8}
\varepsilon^L  = \varepsilon^\prime_0 + \frac{2 e^2}{3\pi}\left(\frac{\beta_0^{-1}}{z^2} - \frac12\,\log\frac{1-\beta}{1+\beta} \right)+ A\,\log \frac{m_{++}m_{+-}}{m_{-+}m_{--}} 
+ \sum_{\lambda=\pm} B_\lambda\,\log \frac{n_{\lambda +}}{n_{\lambda -}} 
\end{equation}
with 
\begin{eqnarray}   \label{b2}
A  & = & \frac{e^2}{6 \pi} \, \left(1- \frac1{2\,\zeta^2}\right)\,\sqrt{1+ \frac1{\zeta^2}} \\[2ex]  \label{b3}
m_{\tau \nu} & = & \nu z + \zeta^2 + \tau\,\beta_0^{-1}\,\zeta\,\sqrt{1+\zeta^2}  \,, \qquad \tau\,,\;\nu =\pm 1
\\[2ex]  \label{b1}
B_\lambda   & = &  \frac{e^2 (\beta_0^{-1}+ \lambda u^\prime) }{12\pi z^3}\,\left[ (\beta_0^{-1}+ \lambda u^\prime)^2 - 3 z^2  \right]      \\[2ex]  \label{b4}
n_{\lambda\sigma} &=& u^\prime(u^\prime+\lambda \beta_0^{-1}) - z(z+\sigma)      \,, \qquad \qquad \lambda,\,\sigma =\pm 1 %\\[2ex]  
\end{eqnarray}
and
\begin{equation} \label{b5}
\varepsilon^\prime_0    = 1 + \frac{e^2}{3\pi}\,\left(\frac53 -\frac1{\zeta^2 } \right)  - 4\,A\,\ln \left(\zeta + \sqrt{1+\zeta^2}\right)   
\end{equation}  

Consider now  (\ref{b8}) as a function of the complex variables $z$ and $u$. The possible singularities are 
\begin{description}
\item[a pole] at $z=0$, which is a false singularity as is revealed by the Taylor expansion (Mathematica)
\begin{eqnarray*}
 \varepsilon^L &=& 1 + \frac{5e^2}{9\pi} + \frac{e^2\,(\mu_0-\beta_0^{-1})}{3\pi \,u^2}  - \frac{2e^2 }{3\pi }\,\ln\frac{1+\beta_0^{-1}}{\mu_0} +  \frac{e^2\,(\mu_0^2 + 2 u^2) \sqrt{u^2 -\mu_0^2}}{6\pi \,u^3}\, \times 
\\[2ex] 
  & &  \left[- 2\,\ln\frac{\sqrt{-u^2} + \sqrt{-u^2+\mu_0^2}}{\mu_0} + \ln\frac{u - \beta_0^{-1} \sqrt{u^2-\mu_0^2}}{u + \beta_0^{-1} \sqrt{u^2-\mu_0^2}} \right]    + O(z^2) 
\end{eqnarray*}
This has a singularity at $u=0$ but, as $z=0$, this implies that there is a singularity at $u=z=0$. For the sake of convenience, we introduce the variables
$$ \kappa_\pm = u \pm z = \frac1{2 q_0}\,\left(\omega \pm q \right) \,\qquad {\rm or} \qquad \kappa_\pm= \frac1{2 q_0}\,k_\pm\,, $$
where $k_\pm$ are the variables (\ref{e5a}), and the pole is at $k_+=k_-=0\,$, which is outside the region $\,\imp(k_\pm)>0\,$. Recall that this is the region $\,\imp(\omega)> |\imp(q)|\,$ ---see Section \ref{S1.2.2}, eq. (\ref{e11f}). 

\item[a pole (or branch point)] at $\, \zeta^2 = - \kappa_+ \kappa_- /\mu_0^2=0\,$, which is outside the region $\,\imp(k_\pm)>0\,$. 

\item[branch points at $\,n_{\lambda\sigma} = 0 $]. Introducing now $\; \beta_0^{-1}=\mu_0 \,\cosh \zeta\;$ and $\;1=\mu_0 \,\sinh \zeta\,$, we have that 
$$ n_{\lambda\sigma}= \left(\kappa_+ + \frac{\lambda \mu_0}{2}\, e^{-\lambda\sigma \zeta}\right)\,\left(\kappa_- + \frac{\lambda \mu_0}{2}\, e^{\lambda\sigma \zeta}\right) -\frac{\mu_0^2}{4} \,. $$
In terms of the new variables 
$$\,a + i \,b := \kappa_+ +  \frac\lambda2\,\mu_0\,e^{-\sigma\lambda\zeta}  \,\qquad {\rm  and} \qquad c + i\,d := \kappa_- + \frac\lambda2\,\mu_0\,e^{\sigma\lambda\zeta}  \,,$$ 
the forbidden region (\ref{e11f}), $\,\imp \kappa_\pm > 0\,$, is $\; b > 0\,$ and $\, d> 0\,$, whereas the branch point equation $n_{\lambda \sigma }=0\,$ becomes   
$$ a c - b d = \frac14\,\mu_0^2 \,, \qquad \qquad a d + b c = 0 \,.$$
This implies that 
$$ a = - c\,\frac{b}{d} \qquad {\rm and} \qquad   - c^2\,\frac{b}{d} - bd = \frac14\,\mu_0^2 \,,$$
which is impossible if $\, b> 0\,$ and $\,d>0\,$ and therefore these branch points lie out of the forbidden region. 

\item[branch points at $\,m_{\tau\nu}  = 0\,$.] Each of these coincides with a $n_{\lambda\sigma}$ for some $\lambda$ and $\sigma$. Indeed, it can be easily checked that $\,\displaystyle{\prod_{\tau,\nu=\pm} m_{\tau,\nu} = \mu_0^4\,\prod_{\lambda,\sigma=\pm} n_{\lambda\sigma}   }\,$, whence it follows that one of the $m$'s vanishes if and only if one of the $n$'s does.

\item[branch points] at $\,\zeta^2 + 1 = 0\,$, which amounts to $\, z^2+\mu_0^2 - u^2 = 0\,$ or
 $\, \kappa_+ \kappa_- = \mu_0^2\,$, which has no roots in the region $\,\imp(k_\pm)>0\,$.  
\end{description}

As for the transverse function, equations (A.4) and (A.4') in ref. \cite{Jancovici1962} written in terms of the dimensionless variables (\ref{ap3aa}) lead to
\begin{equation}   \label{b9}
\varepsilon^T  = \varepsilon^L -\frac{\beta_0^{-1}}{\pi}\,\left(\frac1{z^2} - \frac1{2\mu^2} + \sum_{\lambda=\pm} F_\lambda \,\log \frac{n_{\lambda +}}{n_{\lambda -}} \right)
\end{equation}
with
\begin{equation}   \label{b9a}
F_\lambda = - \frac{1+\lambda \beta_0 u}{8 z}\,\left[1 + \frac1{\zeta^2} -\beta_0^{-2}\,\frac{(1 +\lambda \beta_0 u)^2}{z^2}  \right]
\end{equation}

As seen before, the singularities of $\,n_{\lambda \sigma} = 0\,$ and of $\,F_\lambda\,$, i. e. $\,\mu^2= 0\,$, lie outside the region $\,\imp(u_\pm) = 0\,$, and the Taylor expansion around $z = 0$ yields
$$ \varepsilon^T  - \varepsilon^L = -\frac{e^2 \beta_0^{-1}}{\pi (u^2-\beta_0^{-2})} + O(z^2)  $$
which presents no singularities in the region $\,\imp(k_\pm)>0\,$.

If we take a look at the asymptotic behavior of $\varepsilon^L$ and $\varepsilon^T$ for large $\displaystyle{k_\pm = \omega \pm q}\,$, we see that it is dominated by the logarithmic term in (\ref{b5}) and 
$$ \varepsilon_0^L \sim -\frac{e^2}{3 \pi}\,\log (k_+ k_-) $$ 
which disagrees with the asymptotic behavior (\ref{e12ac}), whence we can conclude that the continuity hypothesis of the derivatives of $G(u^+,u^-)$ are not fulfilled and Theorem \ref{t1} does not apply.

%%%%%%%%%%%%%%%%%%%%%%%%%%%%%%%%%%%%%%%%%%%%%%%%%%

\section{Comparison with other extensions of K-K relations \label{S6}}
\subsection{Leontovich's formula \label{SS6.1}}
As commented before, Leontovitch \cite{Leontovich1961} derived a relation similar to (\ref{e16m}) for the conductivity $\,\sigma(q,\omega)\,$ as a generalization of Kramers-Kr\"onig relations on the basis that: (a) the current density, the electric field and the conductivity $\,\sigma(\omega, q)\,$ are connected similarly as our equation (\ref{e2}), (b) $\,\sigma\,$ is Lorentz invariant and (c) it must fulfill Kramers-Kr\"onig relations for fixed $\,q\,$ in any Lorentzian frame. Furthermore, its derivation considers only one-dimensional space.

Equation (10) in \cite{Leontovich1961} in our notation reads
\begin{equation} \label{MS0}
 \sigma(q,\omega) = \frac1{i \pi}\,\mathcal{P}\int_{\mathbb{R}} \frac{\D \tau}{\tau - \omega}\,\sigma\left(q + \beta [\tau - \omega], \tau  \right)  \,, \qquad \forall |\beta| < 1 \,
\end{equation}
and then Leontovich says without justification that ``it is sufficient that $\sigma(q,\omega)$ satisfy this condition for $\beta=1\,$ \ldots'', that is our relation (\ref{e16m}).  

It is worth noticing that in our approach space has three dimensions (plus isotropy) and we suppose that the response function is relativistically causal but no assumption is made on the Lorentz invariance of $\sigma$.  
In our view, assuming the latter is excessive because the relation (\ref{e1}) connecting $\mathbf{P}$, $\chi$ and $\mathbf{E}$ ---or its analogue relating $\mathbf{j}$, $\sigma$ and $\mathbf{E}$ in Leontovich's approach--- holds in the rest system of the medium only. 

Similarly as in the KK case, in the split form (\ref{e16y}-\ref{e16ya}) we have two relations connecting the real and imaginary parts of $\chi(q,\omega)\,$ ---or $\,\sigma(q,\omega)\,$ in the context of ref. \cite{Leontovich1961}--- but these relations do not look like a Hilbert transform (which is anti-involutive). Perhaps this leads Leontovich to conclude that ``In contrast to [the standard Kramers-Kr\"onig relation], \ldots [the relativistic relation he had derived] not only relates the real and imaginary parts of $\,\sigma(q,\omega)\,$, but also imposes restrictions on each of them separately''.
It is not clear what restriction he is referring to or what its specific form is.
In any case, we have shown that conditions  (\ref{e16y}-\ref{e16ya}) are indeed a concealed Hilbert transform and, moreover, our present approach allows to determine that the constraint is equation (\ref{e16p}).

Later on Sun et al. \cite{SunPuri1989} derived a similar generalization of Kramers-Kr\"onig relations on the basis of signal propagation at a finite speed $v<c\,$ for one space dimension only. They realised that the covariance of the response function ---the conductivity $\sigma$ in that case--- is irrelevant  and that what really matters is causality. Their relation is a double integral ---eq. (13) in \cite{SunPuri1989}--- that is actually our (\ref{kk4a}) and, as commented in Section \ref{S4}, is equivalent to Leontovich's single integral expression. 
Besides, in contrast with our Laplace transform based approach, Sun's derivation cannot conclude anything about the analyticity of $\sigma(q,\omega) \,$ on any domain of $\mathbb{C}^2\,$. 

\subsection{The Melrose-Stoneham approach \label{SS6.2}}
Melrose and Stoneham \cite{Melrose1977} propose a generalization of Kramers-Kr\"onig relations for the polarization tensor $\alpha^{\mu\nu}(\omega,\mathbf{k})$ on the basis that its inverse Fourier transform $\tilde\alpha^{\mu\nu}(t,\mathbf{x})$ vanishes if $(t,\mathbf{x})$ is in the past of the origen for some inertial reference frame, that is
$\tilde\alpha^{\mu\nu}(t,\mathbf{x})$ vanishes whenever $ t- \mathbf{v}\cdot\mathbf{x} < 0 \,, \,  \forall |\mathbf{v}| < 1\,$, 
and therefore
\begin{equation} \label{MS2}
\tilde\alpha^{\mu\nu}(t,\mathbf{x}) = \theta_{\mathbf{v}}(t,\mathbf{x}) \cdot \tilde\alpha^{\mu\nu}(t,\mathbf{x}) \,, \quad  {\rm if} \quad  |\mathbf{v}| < 1\,,\quad {\rm where}\quad \theta_{\mathbf{v}}(t,\mathbf{x}):=\theta(t - \mathbf{v}\cdot\mathbf{x}) \,,
\end{equation} 
The Kramers-Kr\"onig generalization they propose is the Fourier transform of the above relation, namely
\begin{equation} \label{MS3}
\alpha^{\mu\nu}(\omega,\mathbf{k}) = \left(Q_{\mathbf{v}}\ast \alpha^{\mu\nu}\right)(\omega,\mathbf{k}) \,, \quad  \forall |\mathbf{v}| < 1\,,
\end{equation}
where
\begin{equation} \label{MS4}
Q_{\mathbf{v}}(\omega,\mathbf{k}):=\frac1{(2 \pi)^4}\,\left(\mathcal{F}\theta_{\mathbf{v}}\right)(\omega,\mathbf{k})= \frac{ i \, \delta(\mathbf{k}- \omega \mathbf{v})}{2 \pi (\omega + i\,0)}  \,, 
\end{equation}
and they obtain
\begin{equation} \label{MS5}
\alpha^{\mu\nu}(\omega,\mathbf{k}) = \frac1{\pi\, i}\,\mathcal{P}\int_{\mathbb{R}} \frac{\D\zeta}{\zeta -\omega }\,\alpha^{\mu\nu}(\zeta,\mathbf{k}+ [\zeta-\omega]\mathbf{v})   \,, \qquad  \forall |\mathbf{v}| < 1\,,
\end{equation}
which reproduces the  Kramers-Kr\"onig relation for $\mathbf{v}=0\,$.

In the isotropic case,  $\alpha^{\mu\nu}(\omega,\mathbf{k})=\alpha^{\mu\nu}(\omega,|\mathbf{k}|)\,$, for $|\mathbf{v}|=1\,$ the above equation reduces to Leontovich relation if $\mathbf{v}$ parallel to $\mathbf{k}\,$ and to Dolgov relation \cite{Kirzhnits1987,Dolgov1982} if $\mathbf{v}\cdot\mathbf{k}=0\,$.

However, equation ( \ref{MS5}) seems more restrictive than those two particular cases. Indeed, the Melrose generalization is a 3-parameter infinite class of relations that can be refined a bit more if we include the fact that $\alpha^{\mu\nu}$ is the Fourier transform of a function $\tilde\alpha^{\mu\nu}$ vanishing in a large domain. Finer tuning comes from realizing that
\begin{equation} \label{MS6}
 t- \mathbf{v}\cdot\mathbf{x} \geq 0 \,, \quad  \forall |\mathbf{v}| < 1\,, \quad\mbox{if, and only if,} \quad t- |\mathbf{x}| \geq 0  \,,
\end{equation}
whence it follows that
\begin{equation} \label{MS7}
\tilde\alpha^{\mu\nu} = \theta_{\mathbf{v}}\cdot \tilde\alpha^{\mu\nu}\,, \quad  \forall |\mathbf{v}| < 1\,, \quad\mbox{if, and only if,} \quad 
\tilde\alpha^{\mu\nu} = \theta_f \cdot \tilde\alpha^{\mu\nu}\,, \quad {\rm where} \quad  \theta_f(t,\mathbf{x}) := \theta(t-|\mathbf{x}|) \,.
\end{equation}
Defining $Q_f := (2 \pi)^{-4}\,\mathcal{F} \theta_f\,$, we have that 
\begin{equation} \label{MS8}
 Q_{f}(\omega,\mathbf{k}) = \frac1{2 \pi^3 \left[(\omega + i 0)^2 - k^2\right]^2}\,, \qquad k=|\mathbf{k}| \,,
\end{equation}
and, by the convolution theorem, the statement (\ref{MS7}) is equivalent to
\begin{equation} \label{MS9}
\alpha^{\mu\nu} = Q_{\mathbf{v}} \ast \alpha^{\mu\nu}\,, \quad  \forall |\mathbf{v}| < 1\,, \quad\mbox{if, and only if,} \quad 
\alpha^{\mu\nu} = Q_f \ast \alpha^{\mu\nu}  \,.
\end{equation}
Thus the whole class of relations (\ref{MS3}) can be summed up in only one, $\,\alpha^{\mu\nu} = Q_f \ast \alpha^{\mu\nu} \,$, that is
$$ \alpha^{\mu\nu}(\omega,\mathbf{k}) = \frac1{2 \pi^3}\,\int_{\mathbb{R}} \D\zeta \,\int_{\mathbb{R}^3}\,\D \mathbf{q}\,\frac{\alpha^{\mu\nu}(\omega-\zeta,\mathbf{q})}{\left[(\zeta+i0 )^2 - |\mathbf{k}-\mathbf{q}|^2 \right]^2 } \,.$$

So far no assumption has been made about the dependence on the direction of $\mathbf{k}\,$.
In the isotropic case, $\alpha^{\mu\nu}(\omega,\mathbf{k})=\alpha^{\mu\nu}(\omega,k)\,$, using spherical coordinates with $\mathbf{k}$ as the polar axis, we have that
$$ \alpha^{\mu\nu}(\omega,k) = \frac1{\pi^2}\,\int_{\mathbb{R}} \D\zeta \,\int_0^\infty\,\D q\,\frac{2 \,q^2\,\alpha^{\mu\nu}(\omega-\zeta,q)}{\left[(\zeta+i 0 )^2 -k^2 - q^2\right]^2 - 4 \,k^2 q^2} \,   $$
and, with a little algebra and including the fact that $\alpha^{\mu\nu}(\omega,k)$ is even in $k$, it becomes 
\begin{equation} \label{MS10}
\alpha^{\mu\nu}(\omega,k) = - \frac1{2 k \pi^2}\,\int_{\mathbb{R}} \D\zeta \,\int_{\mathbb{R}}\,\D q\,\frac{q\,\alpha^{\mu\nu}(\omega-\zeta,q)}{(\zeta+i0 )^2 - (k - q)^2} \,.  
\end{equation}
For non-dispersive $\alpha^{\mu\nu}(\omega)\,$, the integral on $q$ can be performed and, after a little algebra, it leads to the conventional Kramers-Kr\"onig relation (\ref{e17}).

To compare the above relation with (\ref{kk3}) ---replacing $\chi$ with $\alpha$, of course--- we first introduce the variables $\,\rho=\zeta-k+q\,$ and $\,\tau =\zeta+k-q\,$ to obtain
$$ \alpha^{\mu\nu}(\omega,k) = - \frac1{4 \pi^2}\,\int_{\mathbb{R}^2} \frac{\D\rho \,\D \tau}{(\rho+i 0)\,(\tau+i 0)}\,\left(1 +\frac{\rho-\tau}{2 k}\right)\,\alpha^{\mu\nu}\left(\omega-\frac{\tau+\rho}2,\,k + \frac{\tau-\rho}2 \right) \,  $$
that, in terms of $\,\xi^{\mu\nu}(k_+,k_-) := \alpha^{\mu\nu}(\omega,k)\,$, becomes
$$ \xi^{\mu\nu}(k_+,k_-) = - \frac1{4 \pi^2}\,\int_{\mathbb{R}^2} \frac{\D\rho \,\D \tau}{(\rho+i 0)\,(\tau+i 0)}\,\left(1 +\frac{\rho-\tau}{2 k}\right)\,\xi^{\mu\nu}(k_+-\rho, k_- - \tau)  \,  $$
or
$$ \xi^{\mu\nu}(k_+,k_-) = - \frac1{4 \pi^2}\,\int_{\mathbb{R}^2} \frac{\D\rho \,\D \tau}{(k_+-\rho+i 0)\,(k_- - \tau+i 0)}\,\left(1 +\frac{\rho-\tau}{2 k}\right)\,\xi^{\mu\nu}(\rho,\tau)  \,,  $$
which is the same as relation (\ref{kk3}) because the term with the factor $\,\rho-\tau\,$ in the integral on the right hand side vanishes. Indeed, the factor $\,\rho-\tau\,$ is antisymmetric by interchanging $\rho $ and $\tau$ whereas, due to the symmetry (\ref{e8a}), the other factors in the integral are symmetric. 

Thus, our generalization (\ref{kk3}) of the Kramers-Kronig relations is equivalent to the M-S proposal and at the same time is simpler ---an integral relation in front of a triple infinity.

%%%%%%%%%%%%%%%%%%%%%%%%%%%%%%%%%%%%%%%%%%%%%%%%%%%

\section{Conclusion}
We have studied the consequences of requiring causality on the general form of the susceptibility function $\chi(q,\omega)$ of an isotropic medium in the linear response approximation (or, alternatively, the dielectric function $\varepsilon$). Due to the fact that we are beyond the optical approximation ---$\chi$ depends both on wave vector  and frequency--- the electric polarization, the {\em effect}, at one point $\mathbf{x}$ and a given instant of time $t$ depends on the values of the electric field, the {\em causes}, at space points other than $\mathbf{x}$. Hence the response of the medium involves {\em signals} propagating from the {\em causes} to the {\em effect} and causality requires that these signals do not travel faster than light in vacuum. 

On the basis of merely this causality condition and the requirement that a constant finite electric field must produce a finite polarization we have proved that the extension of the dielectric function to complex values of the variables $q$ and $\omega$  must be double holomorphic in the region $ \,\left|\imp(q)\right| < \imp(\omega)\,$. 

Furthermore, by applying asymptotic theorems for Laplace integrals, we have studied the behavior of  $\,\chi\,$ for large values of $\, \omega \pm q\,$. This requires some additional assumption concerning the continuity of electric  susceptibility $\tilde\chi(\mathbf{r},t)$ in spacetime representation. Similarly as ref. \cite{Jackson1} does, in Section \ref{S1.2.1} we have assumed that $\tilde\chi(\mathbf{r},t)$ is continuous and does not start abruptly at the boundary $ \,t = |\mathbf{r}|/c\,$. However, later on (Section \ref{S5.1}) we have seen that this assumption is not met by Lindhard's semiclassical degenerate electron gas, which is simple enough to allow the Fourier transform to be calculated directly.

We have then applied the Cauchy integral formula to derive the extension of Kramers-Kronig relations to this kind of causality at a distance.
This requires, like in the standard case, to assume that $\,q\,\chi(q,\omega)\,$ decays fast enough for large values of $\,|\omega \pm q|\,$.
We have reobtained a generalization of K-K relations first obtained by Leontovich \cite{Leontovich1961} for conductivity on the basis of Lorentz invariance in 1+1 spacetime dimensions. As a matter of fact we improve Leontovich's result in that: (a) our derivation relies on more general assumptions and (b) we find the concrete form of the extra conditions that Leontovich anounces without proving or specifying. 

The analyticity conditions here obtained are to be taken as a quality test to be passed by any proposal of dielectric function, either empirical or derived from a microscopic model, in order to become acceptable on the basis of causality. Other results, e. g. the asymptotic behavior of $\,\chi(q,\omega)\,$ are less robust since their derivation is based on additional less general assumptions, e. g. the continuity of $\tilde\chi(\mathbf{r},t)$ and its non-abrupt switch on at the boundary $\,t= |\mathbf{r}|\,$.

%%%%%%%%%%%%%%%%%%%%%%%
Often, beyond the optical approximation and to study the causal behavior of a dielectric function $\varepsilon(q,\omega)$, Kramers-Kr\"onig relations have been checked for $q$ real and constant, whereas $\omega$ is a real variable. In that case, as $\imp(q)=0\,$, our condition (\ref{e11g}) implies that $\varepsilon(q,\omega)$ is analytic for $\,\imp(\omega) > 0\,$. Together with a convenient behavior at $|\omega| \rightarrow\infty\,$, the latter implies that the usual Kramers-Kr\"onig relations are fulfilled. However the converse is not true: the compliance with Kramers-Kr\"onig conditions for real values of $q$ does not imply the FS-causality condition (\ref{e11g}). 
%%%%%%%%%%%%%%%%%%%%%%%%%%%%

As an application we have tested several dielectric functions that are found in the literature for different microscopic models of a degenerate electron gas, whether relativistic or not. 
Our intention is not assessing the goodness of these microscopic models but merely seeing the FS-causality condition in operation. 
Whereas the relativistic models have already a {\em built in} FS-causality and they have succesfully passed the analyticity test, in the non-relativistic cases the success depends on the parameters of the model, e. g. if the Fermi momentum exceeds the electron mass time $c$, then the susceptibility function does not pass the analyticity test. 
%%%%%%%%%%%%%%%%%%%%%%%%%%%%

We have compared our results with other proposals in the literature, particularli (a) the one by Leontovich \cite{Leontovich1961} for a one-dimensional space and also (b) those by Melrose-Stoneham \cite{Melrose1977} or (c) by Dolgov-Kirzhnitz \cite{Kirzhnits1987,Dolgov1982} for three space dimensions. 
We have found that (b) is in fact a 3-parameter infinity of conditions, subject to a vector $\,|\mathbf{v}| \leq 1\,$, while (a) and (c) are particular cases of (b) for $\,|\mathbf{v}| = 1\,$ and according to whether $\mathbf{v}$ is parallel or perpendicular to $\mathbf{q}$ respectively.
Finally we have shown that the infinity of conditions (b) are equivalent to our unique condition (\ref{kk4}). 

Although in the body of the work we have adopted the assumptions of isotropy and homothecy (\ref{0a}), in Appendix A we have shown that similar results also hold if the latter assumption is replaced with general linearity.

Our reasoning here is based on the assumption of causality in the relations (\ref{0a}) and (\ref{e2}), which connect an input and an oputput through a response function. Relations of this kind are the basis of linear response theory. Therefore the results here derived also hold for any homogeneous, isotropic response function in the framework of any linear response theory.

%%%%%%%%%%%%%%%%%%%%%%%%%%%%%%%%%%%%%%%%%%
\section*{Acknowledgment}
Funding for this work was partially provided by the Spanish MINCIU and ERDF (project ref. PID2021-123879OB-C22).

\section*{Appendix}
\subsection*{A: Isotropic media}
The relation connecting polarization and electric field is more complex than mere homothecy, such as (\ref{0a}) or (\ref{e2}), and it is rather of matrix type, namely
\begin{equation}  \label{B1}
 P_j(\mathbf{q},\omega) = \chi_{jl}(\mathbf{q},\omega)\,E_l(\mathbf{q},\omega) \,, 
\end{equation}
where summation over repeated indexes is understood. 
For isotropic media the susceptibility matrix has the form
\begin{equation}  \label{B2}
 \chi_{jl}(\mathbf{q},\omega) = \chi^T(q,\omega)\,\left(\delta_{jl} - \hat q_j \hat q_l \right) +  \chi^L(q,\omega)\,\hat q_j \hat q_l \,, 
\end{equation}
with $ \,q= |\mathbf{q}|\,$,  $\,\mathbf{q} = q\,\hat{\mathbf{q}}\,$,  and the scalars $\chi^T$ and $\chi^L$ stand for the transverse and longitudinal susceptibilities respectively. It is obvious that
\begin{equation}  \label{B3}
 \chi_{jj}= 2 \chi^T +  \chi^L \qquad {\rm and } \qquad \chi^L = \chi_{jl}\,\hat q_j \hat q_l   \,.
\end{equation}
%%%%%%%%%%%%%%%%%%%%%%%%%%%%%%%
Our aim is to check whether requiring FS-causality on the relation (\ref{B1}) implies restrictions similar to those derived from the simplest assumptions (\ref{0a}) or (\ref{e2}).

%%%%%%%%%%%%%%%%%%%%%%%%%%%%%%%%

In terms of spacetime variables expression (\ref{B1}) becomes the convolution
\begin{equation}  \label{B4}
 \tilde P_j(\mathbf{x},t) = (2 \pi)^{-2}\,\left(\tilde\chi_{jl} \ast \,\tilde E_l\right)_{(\mathbf{x},t)} \,, 
\end{equation}
where $\tilde\chi_{jl}(\mathbf{x},t)$ is the Fourier transform (\ref{e3}) of $\chi_{jl}(\mathbf{q},\omega)\,$.
Similarly as in (\ref{B2}), in the isotropic case the matrix  $\tilde\chi_{jl}(\mathbf{x},t)$ is
\begin{equation}  \label{B5}
 \tilde\chi_{jl}(\mathbf{x},t) = \alpha(r,t)\,\left(\delta_{jl} - \hat x_j \hat x_l \right) + \beta(r,t)\,\hat x_j \hat x_l \,, 
\end{equation}
with $ \,r= |\mathbf{x}|\,$,  $\,\mathbf{x} = r\,\hat{\mathbf{x}}\,$  and, similarly as in (\ref{B3}),
\begin{equation}  \label{B6}
\tilde \chi_{jj}= 2 \alpha +  \beta \qquad {\rm and} \qquad \beta = \tilde\chi_{jl}\,\hat x_j \hat x_l   \,.
\end{equation}

FS-causality then implies that 
$\quad  \tilde\chi_{jl}(\mathbf{x},t) = 0 \,, \quad{\rm if} \quad |\mathbf{x}| > t\;\,$
and, combining (\ref{B1}-\ref{B6}), after a simple but tedious algebra, we arrive at
%\begin{eqnarray}  \label{B7a}
\begin{equation}  \label{B7a}
q\,\chi^J(q, \omega) = \frac{i}{2 \pi}\,\int_{|r|\leq t} \hspace*{-1em} \D r\,\D t\,e^{i(\omega t - q r)} \,r\, B^J(q,r,t) \,, \qquad J=L,\,T  \,,% \\[2ex]  \label{B7b}
%q\,\chi^T(q, \omega) &=& \frac{i}{2 \pi}\,\int_{|r|\leq t} \hspace*{-1em} \D r\,\D t\,e^{i(\omega t - q r)} \,r\, A(q,r,t) \,, 
\end{equation}
where 
\begin{eqnarray}  \label{B8a}
B^L(q,r,t) &:=& \beta(r,t) - \left(\frac{3 i}{q r} + \frac2{q^2 r^2} \right)\,\left[\beta(r,t) - \alpha(r,t) \right] \,, \\[2ex]  \label{B8b}
B^T(q,r,t) &:=& \alpha(r,t) + \left(\frac{3 i}{2 q r} + \frac1{q^2 r^2} \right)\,\left[\beta(r,t) - \alpha(r,t) \right] \,
\end{eqnarray}
and the extensions $\alpha(r,t) = \alpha(|r|,t)$ and $\beta(r,t) = \beta(|r|,t)\,$ are understood.

A particularly simple combination is the trace
$$ q\,\chi_{jj}(q, \omega) = \frac{i}{2 \pi}\,\int_{|r|\leq t} \hspace*{-1em} \D r\,\D t\,e^{i(\omega t - q r)} \,r\, \left[2\,\alpha(r,t) + \beta(r,t)  \right] \,, $$
which is similar to (\ref{e5}) and therefore the analogous of (\ref{e11g}) also holds for $q\,\chi_{jj}(q, \omega)\,$, i. e. \\[1ex]
\centerline{all singularities of $\;q\,\chi_{jj}(q,\omega)\;$  are in the region $\;\imp(\omega) \leq \left|\imp(q)\right| \,$.  }

\medskip\noindent
Furthermore, in view of equations (\ref{B7a}-\ref{B8b}), the same conclusion can be drawn for $\,q\,\chi^L(q,\omega)\,$ and $\,q\,\chi^T(q,\omega)\,$, perhaps with an additional pole at $q=0\,$.

\subsection*{B: Asymptotic expansions}
To determine the asymptotic behavior of the double Laplace integral
$$  g(s_+,s_-) = \int_{\mathbb{R}^{+ 2}} \D u^+ \,\D u^-\, G(u^+,u^-) e^{-s_+u^+ - s_- u^-} $$
for large values of $\,s_\pm\,$ we shall prove {\bf Theorem \ref{t1}}: 

%%%%%%%%%%%%%%%%%%%%%%%%% \paragraph{Theorem} 
\begin{quote} 
{\em Let $\,G(u^+,u^-)\,$ have continuous partial derivatives $\,G^{(l,j)}(u^+,u^-) = \partial_+^l \partial_-^j G(u^+,u^-)\,$ up to the $N$-th order, for $0\leq u^\pm \leq a^\pm\,$, and assume that there exist $\,a_\pm \in \mathbb{R} \,$ such that
$$\,\int_{\mathbb{R}^{+ 2}} \D u^+ \,\D u^-\, G^{(l,j)}(u^+,u^-) e^{-a_+u^+ - a_- u^-} < \infty\,, \qquad \quad l + j \leq N  \,,$$
then 
\begin{equation}  \label{A0}  
 g(\lambda s_+,\lambda s_-) = \sum_{0\leq l+j\leq N-1} \frac{G^{(l,j)}(0^+,0^+)}{\lambda^{l+j+2} s_+^{l+1} s_-^{j+1}} + o(\lambda^{-N-1}) \,, \qquad \qquad \forall \lambda \in \mathbb{R}^+ \,.
\end{equation} 
}
\end{quote}
%%%%%%%%%%%%%%%%%%%%%%%%%%

\noindent
\paragraph{Proof:} We will take advantage of the next result \cite{Erdelyi} on the asymptotic expansion of Laplace integrals:
\paragraph{Lemma.} {\em Let $\;f(x) = \int_0^\infty e^{-x t} \phi(t)\,\D t\,$. If $\phi(t)$ is $N $ times continuously differentiable for $0 \leq t \leq a\,$ and, for some $x_0\,$, $\;f(x_0) < \infty\,$, then
\begin{equation}  \label{A1}
f(x) = \sum_{n=0}^{N-1} \frac{\phi^{(n)}(0)}{x^{n+1}} + o\left(x^{-N-1}\right) \,,
\end{equation}
uniformly in $\,{\rm arg}\,x\,$, as $ x \rightarrow\infty\,$, in $\,\displaystyle{\left|{\rm arg}\,x\right| < \frac{\pi}2 - \Delta\,, \quad \Delta < 0}\,$.}

The symbol ``$o$'' is to be understood as 
$$ g(x) = o(h(x)) \,, \qquad {\rm for} \quad  x \rightarrow\alpha\,, \qquad {\rm if} \qquad 
\lim_{x \rightarrow\alpha} \frac{g(x)}{h(x)} = 0 \,. $$

We start writing $\;\displaystyle{g(s_+,s_-) = \int_0^\infty \D u^+ \, e^{-s_+u^+}\,\Gamma(u^+,s_-) }\,$, with
\begin{equation}  \label{A2}
\Gamma(u^+,s_-) :=  \int_0^\infty \D u^-\, e^{- s_- u^-} \, G(u^+,u^-) \,.
\end{equation}
Applying the Lemma, we obtain 
\begin{equation}  \label{A3}
g(s_+,s_-) = \sum_{n=0}^{N-1} \frac{\partial^n_+\Gamma(u^+=0,s_-)}{s_+^{n+1}} + o\left(s_+^{-N-1}\right)  \, 
\end{equation}
and, from (\ref{A2}) it follows that
$$ \partial^n_+\Gamma(u^+=0,s_-) = \int_0^\infty \D u^-\, e^{- s_- u^-} \,\partial^n_+ G(u^+=0,u^-) \,, \qquad \left|{\rm arg}\,s_-\right| < \frac{\pi}2 - \Delta_-\,, \quad \Delta_- < 0\, $$
and, since $\,\partial^n_+ G(u^+=0,u^-)\,$ is $N-n$ times continuously differentiable, we can apply the Lemma again to write
$$ \partial^n_+\Gamma(u^+=0,s_-) = \sum_{k=0}^{N-n-1} \frac{\partial_-^k\partial^n_+ G(0,0)}{s_-^{k+1}} + o\left(s_-^{-N+n-1}\right)  \, $$
and, replacing it in (\ref{A3}), we have that
$$ g(s_+,s_-) = \sum_{n=0}^{N-1} \sum_{k=0}^{N-n-1} \frac{\partial_-^k\partial^n_+ G(0,0)}{s_-^{k+1} s_+^{n+1}} + \sum_{n=0}^{N-1} \frac1{s_+^{n+1}}\,o\left(s_-^{-N+n-1}\right) +
o\left(s_+^{-N-1}\right)  \,.$$
 
Finally, as $\,s_+^{-n-1} = o(s_+^ {-n})\,$ and $\,o(g)\cdot o(h) = o(g\cdot h)\,$, we arrive at
$$ g(s_+,s_-) = \sum_{0 \leq k+n \leq N-1} \frac{\partial_-^k\partial^n_+ G(0,0)}{s_-^{k+1} s_+^{n+1}} + \sum_{n=0}^{N-1} o\left(s_+^{-n}s_-^{-N+n-1}\right) +
o\left(s_+^{-N-1}\right)  \,$$
and replacing $\,s_\pm\,$ with $\,\lambda s_\pm\,$ equation (\ref{A0}) follows. \hfill $\Box$

%%%%%%%%%%%%%%%%%%%%%%%%%%%%%%%%%%%%%%%%%%%%%%%%%%%%%%%%%%%%%%%%%%%%%%%%%%

\end{document}